\documentclass[iop,letter]{emulateapj}

\usepackage{natbib}
\usepackage{graphicx}
\usepackage{times}
\usepackage{apjfonts}
\usepackage{color}
\usepackage{url}
\usepackage{hyperref}

\hypersetup{ colorlinks,
  linkcolor=blue,
  filecolor=blue,
  urlcolor=blue,
  citecolor=blue}
\def\ltsima{$\; \buildrel < \over \sim \;$}
\def\simlt{\lower.5ex\hbox{\ltsima}}
\def\gtsima{$\; \buildrel > \over \sim \;$}
\def\simgt{\lower.5ex\hbox{\gtsima}}
\def\simless{\mathbin{\lower 3pt\hbox
   {$\rlap{\raise 5pt\hbox{$\char'074$}}\mathchar$}}}   
\def\simgreat{\mathbin{\lower 3pt\hbox
   {$\rlap{\raise 5pt\hbox{$\char'076$}}\mathchar"7218$}}}   

\def\rosat{{\it ROSAT}}
\def\xmm{{\it XMM-Newton}}
\def\chandra{{\it Chandra}}
\def\J2327{ACT-CL~J2327.4$-$0204}
\def\JZZFF{ACT-CL~J0044.4$+$0113}
\def\lcdm{$\Lambda$CDM}
\def\kms{km~s$^{-1}$}

\slugcomment{Draft 0.1}
\submitted{Submitted to the Astrophysical Journal, October 15, 2012; Accepted, December 22, 2012}

\shorttitle{Physical Properties of ACT Equatorial Clusters}
\shortauthors{Menanteau et al.}

\begin{document}

\title{The Atacama Cosmology Telescope: Physical Properties of
  Sunyaev-Zel'dovich Effect clusters on the Celestial Equator\footnotemark[\dag]\footnotemark[\ddag]}

\author{Felipe Menanteau\altaffilmark{1,*},
Crist\'obal~Sif\'on\altaffilmark{2,3,*},
L.~Felipe~Barrientos\altaffilmark{2},
Nicholas~Battaglia\altaffilmark{4},
J.~Richard~Bond\altaffilmark{5},
Devin~Crichton\altaffilmark{6},
Sudeep~Das\altaffilmark{7},
Mark~J.~Devlin\altaffilmark{8},
Simon~Dicker\altaffilmark{8},
Rolando~D\"unner\altaffilmark{2},
Megan~Gralla\altaffilmark{6},
Amir~Hajian\altaffilmark{5},
Matthew~Hasselfield\altaffilmark{9},
Matt~Hilton\altaffilmark{10},
Adam~D.~Hincks\altaffilmark{5},
John~P.~Hughes\altaffilmark{1,*},
Leopoldo~Infante\altaffilmark{2},
Arthur~Kosowsky\altaffilmark{11},
Tobias~A.~Marriage\altaffilmark{6},
Danica~Marsden\altaffilmark{12},
Kavilan~Moodley\altaffilmark{10},
Michael~D.~Niemack\altaffilmark{13,18},
Michael~R.~Nolta\altaffilmark{5},
Lyman~A.~Page\altaffilmark{14},
Bruce~Partridge\altaffilmark{15},
Erik~D.~Reese\altaffilmark{8},
Benjamin~L.~Schmitt\altaffilmark{8},
Jon~Sievers\altaffilmark{14},
David~N.~Spergel\altaffilmark{16},
Suzanne~T.~Staggs\altaffilmark{14},
Eric~Switzer\altaffilmark{5},
Edward~J.~Wollack\altaffilmark{17}
}

\affil{$^1$Rutgers University, Department of Physics \& Astronomy, 136 Frelinghuysen Rd, Piscataway, NJ 08854, USA }
\affil{$^2$Departamento de Astronom{\'{i}}a y Astrof{\'{i}}sica, Facultad de F{\'{i}}sica, Pontificia Universidad Cat\'{o}lica de   Chile, Casilla 306, Santiago 22, Chile}
\affil{$^3$Leiden Observatory, Leiden University, PO Box 9513, NL-2300 RA Leiden, Netherlands}
\affil{$^4$McWilliams Center for Cosmology, Carnegie Mellon University, Department of Physics, 5000 Forbes Ave., Pittsburgh PA, USA, 15213}
\affil{$^5$Canadian Institute for Theoretical Astrophysics,  University of Toronto, Toronto, ON, Canada M5S 3H8}
\affil{$^6$Department of Physics and Astronomy, The Johns Hopkins University, Baltimore, Maryland 21218-2686, USA}
\affil{$^7$High Energy Physics Division, Argonne National Laboratory, 9700 S Cass Avenue, Lemont, IL 60439}
\affil{$^8$University of Pennsylvania, Physics and Astronomy, 209 South 33rd Street, Philadelphia, PA 19104, USA}
\affil{$^9$Department of Physics and Astronomy, University of British Columbia, Vancouver, BC, Canada V6T 1Z4}
\affil{$^{10}$Astrophysics \& Cosmology Research Unit, School of Mathematics, Statistics \& Computer Science, University of KwaZulu-Natal, Durban, South Africa}
\affil{$^{11}$University of Pittsburgh, Physics \& Astronomy Department, 100 Allen Hall, 3941 O'Hara Street, Pittsburgh, PA 15260, USA}
\affil{$^{12}$Department of Physics, University of California Santa Barbara, CA 93106, USA}
\affil{$^{13}$NIST Quantum Devices Group, 325 Broadway Mailcode 817.03, Boulder, CO, USA 80305}
\affil{$^{14}$Joseph Henry Laboratories of Physics, Jadwin Hall, Princeton University, Princeton, NJ, 08544, USA}
\affil{$^{15}$Department of Physics and Astronomy, Haverford College, Haverford, PA 19041, USA}
\affil{$^{16}$Department of Astrophysical Sciences, Peyton Hall, Princeton University, Princeton, NJ, 08544, USA}
\affil{$^{17}$NASA/Goddard Space Flight Center, Greenbelt, MD 20771, USA}
\affil{$^{18}$Department of Physics, Cornell University, Ithaca, NY, USA 14853}

\footnotetext[\dag]{Based on observations obtained at the Gemini Observatory,
    which is operated by the Association of Universities for Research
    in Astronomy, Inc., under a cooperative agreement with the NSF on
    behalf of the Gemini partnership: the National Science Foundation
    (United States), the Science and Technology Facilities Council
    (United Kingdom), the National Research Council (Canada), CONICYT
    (Chile), the Australian Research Council (Australia),
    Minist\'{e}rio da Ci\^{e}ncia, Tecnologia e Inova\c{c}\~{a}o
    (Brazil) and Ministerio de Ciencia, Tecnolog\'{i}a e
    Innovaci\'{o}n Productiva (Argentina)}

\footnotetext[\ddag]{Based in part on observations obtained with the Apache Point
Observatory 3.5-meter telescope, which is owned and operated by the
Astrophysical Research Consortium.}

\footnotetext[*]{Visiting astronomer, Gemini South Observatory.}

\begin{abstract}

  We present the optical and X-ray properties of 68 galaxy clusters
  selected via the Sunyaev-Zel'dovich Effect at 148\,GHz by the
  Atacama Cosmology Telescope (ACT). Our sample, from an area of 504
  square degrees centered on the celestial equator, is divided into
  two regions.
  The main region uses 270 square degrees of the ACT survey that
  overlaps with the co-added $ugriz$ imaging from the Sloan Digital
  Sky Survey (SDSS) over Stripe 82 plus additional near-infrared
  pointed observations with the Apache Point Observatory 3.5-meter
  telescope.  We confirm a total of 49 clusters to $z\approx1.3$, of
  which 22 (all at $z>0.55$) are new discoveries.
  For the second region the regular-depth SDSS imaging allows us to
  confirm 19 more clusters up to $z\approx0.7$, of which 10 systems
  are new.
  We present the optical richness, photometric redshifts, and
  separation between the SZ position and the brightest cluster galaxy
  (BCG). We find no significant offset between the cluster SZ centroid
  and BCG location and a weak correlation between optical richness and
  SZ-derived mass.
  We also present X-ray fluxes and luminosities from the \rosat\ All
  Sky Survey which confirm that this is a massive sample. 
  One of the newly discovered clusters, \JZZFF\ at $z=1.1$
  (photometric), has an integrated \xmm\ X-ray temperature of
  $kT_X=7.9\pm1.0$~keV and combined mass of $M_{200a} =
  8.2_{-2.5}^{+3.3}\times10^{14}\,h_{70}^{-1}M_\odot$ placing it among
  the most massive and X-ray-hot clusters known at redshifts beyond
  $z=1$.
  We also highlight the optically-rich cluster \J2327\ (RCS2~2327) at
  $z=0.705$ (spectroscopic) as the most significant detection of the
  whole equatorial sample with a \chandra-derived mass of
  $M_{200a}=1.9_{-0.4}^{+0.6}\times10^{15}\,h_{70}^{-1}M_\odot$, in
  the ranks of the most massive known clusters like El Gordo and the
  Bullet Cluster.

\end{abstract}


\section{Introduction}

Clusters of galaxies are the cosmic signposts for the largest
gravitationally bound objects in the Universe. Their formation and
evolution as a function of look-back time provides a measurement of
cosmological parameters that complements those obtained from
observations of the cosmic microwave background \citep[CMB,
e.g.,][]{Komatsu2011,Dunkley2011,Reichardt2012}, Type Ia Supernovae
\citep[e.g.,][]{Hicken2009,Sullivan2011,Suzuki2012} and baryon
acoustic oscillations \citep[e.g.,][]{Percival2010}.
The number of clusters as a function of redshift, as demonstrated by
X-ray and optically-selected samples
\citep[e.g.,][]{Vikhlinin2009,Rozo2010} provides a strong constraint on
both the expansion history of the Universe and the gravitational
growth of structure within it \citep[for a recent review
see][]{Allen2011}.

The hot gas in galaxy clusters leaves an imprint on the CMB radiation
through the Sunyaev-Zel'dovich effect \citep[SZ;][]{SZ1972}. The SZ
effect has a frequency dependence that produces temperature shifts of
the CMB radiation corresponding to a decrement below and an increment
above the ``null'' frequency near 220\,GHz \citep[see][for recent
reviews]{Birkinshaw1999,Carlstrom2002}.

Several experiments are now able to carry out large-area cosmological
surveys using the SZ effect. The Atacama Cosmology Telescope (ACT) and
South Pole Telescope (SPT) are providing samples of galaxy clusters
over hundreds of square degrees at all redshifts
\citep{Stan09,Hincks2010,Menanteau-SZ, Marriage2011, Vanderlinde2010,
  Williamson2011, Reichardt2012}, while the {\em Planck} Satellite
probes the full sky for clusters up to redshifts of $z\approx0.55$
\citep{Planck-SZ}.

Although modest in size, the new SZ cluster samples have proven useful
for constraining cosmological parameters \citep{Vanderlinde2010,
  Sehgal2011, Reichardt2012, Hasselfield2012} and have
opened a new window into the extreme systems, the most massive
clusters at high redshift \citep[e.g.,][]{Foley2011,Menanteau2012},
prompting studies that match their observed numbers with the
abundance predictions of the standard $\Lambda$CDM cosmology
\citep[e.g.,][]{Hoyle2011,Mortonson2011,Waizmann2012}.

ACT is a millimeter-wave, arcminute-resolution telescope
\citep{Fowler2007,Swetz2011} designed to observe the CMB on arcminute
angular scales \citep{Dunner2012}. The initial set of ACT observations
during the 2008 season focused on surveying a 455 deg$^2$ region of
the southern sky centered near declination -55$^\circ$ (hereafter
``the southern strip'').
Our previous work studied 23 high-significance clusters from the southern strip
\citep{Marriage2011} with optical confirmations \citep{Menanteau-SZ}. 
One of the highlights of this previous work was the discovery of the
spectacular El Gordo (ACT-CL J0102$-$4915) cluster merger system
at $z=0.87$ \citep{Menanteau2012}.

During the 2009 and 2010 seasons, ACT mainly surveyed a long, narrow
region of the celestial equator that nearly completely overlaps with
the publicly-available optical co-added images from the Sloan Digital
Sky Survey (SDSS) of Stripe 82 \citep[hereafter S82;][]{Annis2011}.
SDSS provides an immediate optical follow-up of clusters that is of
high quality, uniform and at a depth sufficient to detect massive
clusters to $z\approx1$. This is currently unique for high resolution SZ
experiments. Furthermore, the uniform SDSS coverage of S82 has allowed
combined CMB-optical studies such as the detection of the SZ decrement
from low mass (few $10^{14}$ $M_\odot$) haloes by stacking Luminous
Red Galaxies \citep[LRGs,][]{Hand2011}, the first detection of the
kinematic SZ effect \citep{Hand2012}, and the cross-correlation of the
ACT CMB lensing convergence maps \citep{Das2011b} and quasars
\citep{Sherwin2012}.

While the number density of SZ-selected clusters is a potentially
strong cosmological probe, the confirmation of candidates as true
clusters and the determination of their masses is the first and
most fundamental step. 
In this paper we provide the optical and near-infrared (NIR)
confirmation of SZ cluster candidates from 504 square degrees of the
148\,GHz ACT 2009-2010 maps of the celestial equator. Over the ACT
survey area that overlaps S82 (270\,deg$^2$) we use the $ugriz$
optical images, supplemented with targeted NIR observations, to
identify 49 clusters up to $z\approx1.3$.
For targets outside S82  we use the regular-depth SDSS data
from Data Release 8 \citep[DR8;][]{DR8} to confirm 19 clusters.
The contiguous coverage provided by SDSS allows us to investigate
potential offsets between the clusters optical and SZ centroid
position as well as the relation between optical richness and SZ
signal.
In a companion paper \citep{Hasselfield2012} we present a full
description of the SZ cluster selection technique as well as
cosmological implications using the cluster sample. Recently
\cite{Reese2012} presented high-resolution follow-up observations with
the Sunyaev-Zel'dovich Array (SZA) for a small sub-sample of the
clusters presented here.

Throughout this paper we quote cluster masses as $M_{200a}$ (or
$M_{500c}$) which corresponds to the mass enclosed within a radius
where the overdensity is 200 (500) times the average (critical) matter
density. 
We assume a standard flat \lcdm\ cosmology with $\Omega_m=
0.27$ and $\Omega_\Lambda=0.73$, and $E(z)=\sqrt{\Omega_M
  (1+z)^3+\Omega_\Lambda}$. We give relevant quantities in terms of
the Hubble parameter $H_0 = 70\,h_{70}$~km~s$^{-1}$ Mpc$^{-1}$.
The assumed cosmology has a small effect on the cluster mass, for
example if we assume a canonical scaling of $M\propto E(z)^{-1}$, this
implies a $2\%-3\%$ increase between $\Omega_m=0.27$ and
$\Omega_m=0.30$ for a flat cosmology at $0.4<z<0.8$.
In our analysis we convert masses with respect to average or critical
at different overdensities using scalings derived from a
\citet[][NFW]{NFW} mass profile and the concentration-mass relation,
$c(M,z)$, from simulations \citep{Duffy2008}.
All magnitudes are in the SDSS $ugriz$ AB system and all quoted errors
are 68\% confidence intervals unless otherwise stated.

\section{Observations}

The detection of cluster candidates was performed from
matched-filtered ACT maps at 148\,GHz, while confirmation of the
clusters was done using a combination of optical and near-infrared
(NIR) imaging, and archival \rosat\ X-ray data. In the following
sections we describe the procedure followed.

\subsection{SZ observations}

\begin{figure*}[t]
\centerline{\includegraphics[width=7.4in]{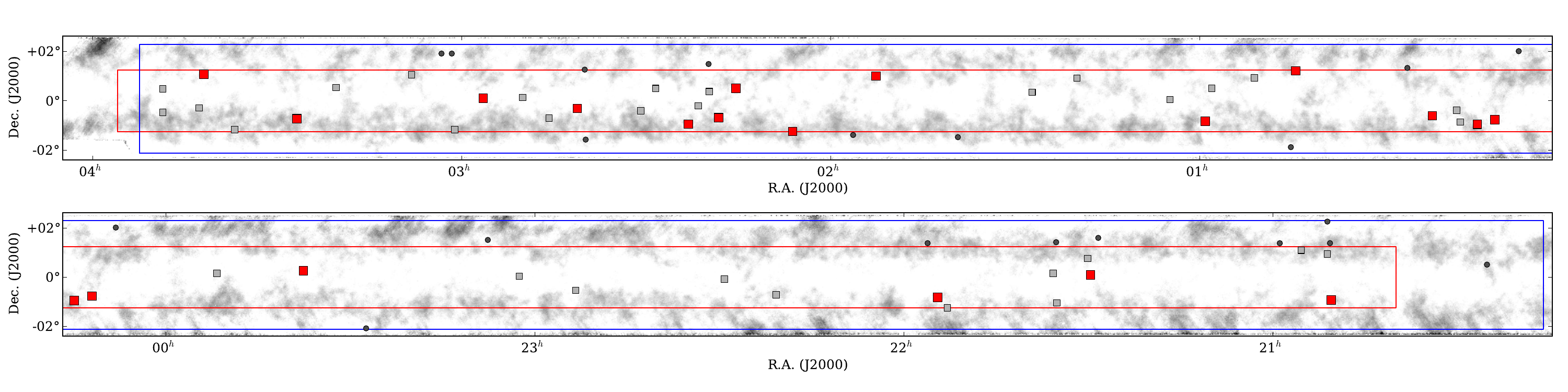}}
\caption{ACT equatorial survey coverage with the SDSS Stripe 82 deep
  optical survey region indicated as red box. The red squares are the
  18 clusters in the pure sample in S82 (see
  Section~\ref{sec:purity}). Gray squares represent the rest of the
  confirmed clusters in the S82 region. Circles are other
  confirmed ACT SZ clusters outside S82. The blue box represents total
  area ($504$~deg$^2$) covered by ACT.}
\label{fig:actmap}
\end{figure*}

ACT operates at three frequency bands centered at 148\,GHz (2.0\,mm),
218\,GHz (1.4\,mm) and 270\,GHz (1.1\,mm), each band having a dedicated
1024-element array of transition-edge-sensing bolometers.
The 270\,GHz band is not as sensitive as the lower frequency channels
and the analysis of it is ongoing although not yet complete.
ACT has concluded four seasons of observations (2007-2010) surveying
two sky areas: the southern strip near declination $-55^\circ$ and a
region over the celestial equator.
In this paper, we apply similar techniques as the ones used on the
southern strip for SZ cluster detection and optical identification
\citep{Menanteau-SZ,Marriage2011} to the ACT 148\,GHz equatorial data.
A full description of the map making procedure from the ACT
time-ordered data is described in \cite{Dunner2012}.


Cluster candidates were detected in the 148\,GHz ACT equatorial maps
over a 504\,deg$^2$ region bounded by $20^h16^m<$RA$<3^h52^m$ and
$-2^\circ07'<$Dec.$<+2^\circ18'$ as shown in Figure~\ref{fig:actmap}
as a blue box over the ACT map \citep{Hasselfield2012}. Nearly fully
contained within this region lies the S82 optical imaging area (shown
as a red box in Figure~\ref{fig:actmap}) which spans
$20^h40^m<$RA$<4^h0^m$ and $-1^\circ15'<$Dec.$<+1^\circ15'$ and covers
275\,deg$^2$.
The effective overlap between the S82 imaging and the ACT maps is
270\,deg$^2$ and corresponds to the deepest section of the ACT data
in the equatorial survey. This constitutes the core of the data we use
in this paper to characterize the SZ selection function.
In the ACT region of the maps beyond the S82 coverage we use the
normal-depth legacy survey from the SDSS DR8. The effective beam for
the 148\,GHz band for the 2009 and 2010 combined seasons has a FWHM of
$1.\!'4$.

Here we highlight the principal aspects of the SZ cluster detection
procedure described in \cite{Hasselfield2012} to provide context
for the characterization of the cluster sample.
After subtracting bright sources from the ACT 148\,GHz source catalog
(corresponding to 1\% of the map area), the map is match-filtered in
the Fourier domain using a set of signal templates based on the
Universal Pressure Profile (UPP) of \cite{Arnaud2010} modeled with a
generalized NFW profile \citep[][Appendix~A]{Nagai2007} as a function
of physical radius.
We use signal templates with FWHM of $0.\!'4$ to $9.\!'2$ in
increments of $0.\!'4$ (23 sets) to match-filter the ACT 148\,GHz maps
to optimize signal-to-noise (S/N) on cluster-shaped objects with an SZ
spectral signature.
Cluster candidates are identified in the filtered maps 
as pixels with S/N$>4$ using the core scale in which
the cluster was most significantly detected. The catalog of cluster
candidates contains positions, central decrements ($\Delta T$), and
the local map noise level. Candidates seen at multiple filter scales
are cross-identified if the detection positions are within $1'$.

\subsection{SDSS Optical Data}

The main optical data set used for the SZ cluster confirmation is the
S82 optical imaging that almost completely overlaps with the deepest
region of the ACT equatorial maps with an effective area coverage of
270\,deg$^2$. 
ACT's survey over S82 is unique for high resolution
SZ experiments, since it provides immediate optical follow-up of an
extremely high and uniform quality at a depth sufficient to detect
massive clusters to $z\approx1$. 
Beyond this common region we use the shallower single-pass data from
DR8 to confidently report cluster identification to $z\approx0.5$.

The S82 survey is a 275\,deg$^2$ stripe (represented by the red box in
Figure~\ref{fig:actmap}) of repeated $ugriz$ imaging centered on the
Celestial Equator in the Southern Galactic Cap, as described in
\cite{Annis2011}. The multi-epoch scanning of the $2^\circ\!\!.5$--wide
SDSS camera provides between 20 to 40 visits for any given section of
the survey which, after aligning and averaging (i.e., co-adding), results
in significantly deeper data.
The co-added S82 images reach $\sim2$ magnitudes deeper than the 
single-pass SDSS data and have a median seeing of $\sim1.\!\!''1$ with a
reported 50\% completeness for galaxies at $r=23.5$ and $i=23$, while
for DR8 this completeness level is reached at $r=21.5$
\citep{Annis2011}. Photometric calibration has typical variation of
$0.5\%$ for $griz$ and $1\%-2\%$ for $u$ across the survey. In
Figure~\ref{fig:mr_z} we show the detection limits for the S82 and DR8
photometry as compared to the observed magnitudes of early-type
galaxies of different luminosities at different redshifts.

The co-added, photometrically-calibrated images and catalogs for S82
were released in October 2008 as part of the SDSS Data Release 7
\citep[DR7;][]{DR7} and are available at the SDSS Data Archive Server
(DAS)\footnote{http://das.sdss.org} and the Catalog Archive Server
(CAS),\footnote{http://casjobs.sdss.org/casjobs} respectively. The
co-added data were run through the SDSS pipelines; the standard SDSS
flag set is available for all objects.

We retrieved Galactic-extinction-corrected {\tt modelMag} photometry
in all 5 bands for all galaxies from the PhotoObj table designated
from runs 106 and 206 under the CAS {\tt Stripe82} database to create
galaxy catalogs, which we split in $0^h20^m$ wide tiles in right
ascension with no overlap between them to avoid object duplication. As
the {\tt Stripe82} database does not include spectroscopic
information, for each galaxy we used the DR8 CAS database for a
spectroscopic redshift, which was ingested into the catalogs if
available \citep{DR8}.
In order to optimize and speed up our cluster identification we fetched
all $ugriz$ fits images for S82 from run numbers 100006 (North) and
200006 (South) and stored them locally to query later. The pixel scale
of the co-added images is $0.\!\!''396$/pixel for all bands.

We compute photometric redshifts for all objects in the S82
photometric catalog using the spectral-energy-distribution (SED) based
Bayesian Photometric Redshift code \citep[BPZ,][]{BPZ} with no
prior. We use the dust-corrected $ugriz$ {\tt modelMag} magnitudes and
the BPZ set of template spectra described in \cite{Benitez-04}, which
in turn is based on the templates from \cite{C80} and \cite{K96}. This
set consists of El, Sbc, Scd, Im, SB3, and SB2 and represents the
typical SEDs of elliptical, early/intermediate-type spiral, late-type
spiral, irregular, and two types of starburst galaxies,
respectively. For the targets with NIR follow-up observations, the
catalogs are augmented by including the $K_S$--band imaging.
The final results are catalogs with photometric redshifts for all
galaxies in S82 augmented by spectroscopic redshifts as available.

For a fraction of the SZ cluster candidates outside the common area
between the ACT equatorial maps and S82, we use regular-depth SDSS
imaging from DR8 to confirm clusters. 
We also retrieved $ugriz$ Galactic-extinction-corrected {\tt
  modelMag} magnitudes for galaxies, but, unlike for the S82, we only
query the DR8 CAS database within a radius of $1^\circ$ of each
candidate. Similarly we only fetched and combined images from tiles
surrounding each candidate to create $10'$ fits images in all 5
bands. Given that the DR8 CAS database provides well-tested
training-set-based photometric redshifts we do not compute our own
SED-based estimates, as we did for S82, and instead we rely on the
ones available in the database. In Section~\ref{sec:redshifts} we
discuss the accuracy of the photometric redshift measurements.

\subsection{Near Infrared Imaging}
\label{sec:APO}

Additional pointed follow-up NIR observations with the Near-Infrared
Camera and Fabry-Perot Spectrometer (NICFPS) on the ARC 3.5-m
telescope of the Apache Point Observatory (APO) aided the confirmation
of five high redshift clusters with S/N$>4.7$. These clusters did not
have a secure optical cluster counterpart in the deep S82 area.
The observations were carried out on UT 2010 Oct 27-28, UT 2011 Nov 02
and UT 2011 Nov 06 when the seeing varied between $0.9
-1.\!''4$. NICFPS is equipped with a $1024\times1024$ Hawaii-I RG
array with $0.\!''273$ pixels and a $4.\!'58$ square field of view. We
obtained between 1800 -- 3870\,s of integration in the $K_S$ band on
each candidate, using 30\,s exposures with 8 Fowler samples per
exposure \citep{Fowler1990}, in a repeating 5 point dither pattern
with box size $20''$. The individual exposures were dark subtracted,
distortion corrected, flat fielded (using a sky flat made from the
science frames), and sky subtracted (using a running median
method). SExtractor \citep{SEx} was used to produce object masks used
in constructing the sky flat and sky images used in the latter two
processing steps. The individual exposures were then astrometrically
calibrated using SCAMP \citep{Bertin2006} and, finally, median
combined using SWARP \citep{Bertin2002}. The photometric zero point
(on the Vega system) for each image was bootstrapped from the
magnitudes of UKIDSS LAS \citep{Lawrence2007} sources in each field
and transformed into the AB system for consistency with the SDSS data.
The estimated uncertainty on the zero point
spans the range 0.01 -- 0.06 mag, with median 0.02 mag. The final
images reach $5\sigma$ depth 18.8 -- 20.2 mag (median 19.4 mag;
measured within a $3''$ diameter aperture), estimated by placing 1000
apertures in each image at random positions where objects are not
detected.
In Table~\ref{tab:APO} we summarize the NIR observations for the
confirmed clusters, which we also discuss in Section~\ref{sec:S82}.

For the clusters with NIR imaging, we registered the $K_S$ and optical
data to create a detection image from the $\chi^2$ quadratic sum
combination of the $i$ and $K_S$--bands using SWARP. Source detection and
photometric catalogs were performed using SExtractor \citep{SEx} in
dual-image mode in which sources were identified on the detection
images using a $1.5\sigma$ detection threshold, while magnitudes were
extracted at matching locations from all other bands.
For clusters with NIR imaging, we use the isophotal magnitudes in the
new catalogs to compute photometric redshifts using the same procedure
described in Section~\ref{sec:redshifts}, with the only variation
being the use of six filters instead of five.

\begin{deluxetable}{lccc} 
\tablecaption{APO NIR observations of Stripe 82 Clusters} 
\tablehead{
\colhead{ACT Descriptor} & 
\colhead{Date Obs.} & 
\colhead{Exp. Time} &
\colhead{photo-z} 
}
\startdata
ACT-CL~J0012.0$-$0046 & UT 2011, Nov 02 & 3870~s & $1.36 \pm 0.06$\\
ACT-CL~J0044.4$+$0113 & UT 2011, Nov 06 & 3600~s & $1.11 \pm 0.03$\\
ACT-CL~J0336.9$-$0110 & UT 2010, Oct 27 & 3600~s & $1.32 \pm 0.05$\\
ACT-CL~J0342.0$+$0105 & UT 2010, Oct 28 & 3150~s & $1.07 \pm 0.06$\\
ACT-CL~J2351.7$+$0009 & UT 2011, Oct 02 & 1800~s & $0.99 \pm 0.03$
\enddata
\label{tab:APO}
\end{deluxetable}
\vspace{0.5cm}

\subsection{\rosat\ X-ray Observations}

We extracted X-ray fluxes for all optically confirmed ACT equatorial
clusters using the \rosat\ All-Sky Survey (RASS) data following the
same procedure as in \citet{Menanteau-Hughes-09} and
\cite{Menanteau-SZ}.  The raw X-ray photon event lists and exposure
maps were downloaded from the MPE {\em ROSAT}
Archive\footnote{ftp://ftp.xray.mpe.mpg.de/rosat/archive/} and were
queried with our own custom software.  At the ACT SZ position of each
cluster, RASS count rates in the $0.5-2$~keV band (corresponding to PI
channels 52--201) were extracted from within radii of 3$^\prime$ for
the source emission and from within a surrounding annulus (5$^\prime$
to 25$^\prime$ inner and outer radii) for the background emission.
The background-subtracted count rates were converted to X-ray
luminosity (in the 0.1--2.4 keV band) assuming a thermal spectrum
($kT_X=5$~keV) and the Galactic column density of neutral hydrogen ($N_H$)
appropriate to the source position, using data from the
Leiden/Argentine/HI Bonn survey \citep{kalberla05}. In
Tables~\ref{tab:S82RASS} and \ref{tab:DR8RASS} we show the X-ray fluxes
and luminosities for all ACT clusters, regardless of the significance
of the RASS detection. Uncertainties are estimated from the count rates
and represent statistical errors.

\section{Analysis and Results}

Our analysis provides a sample of optically-confirmed SZ clusters from
the ACT cluster candidates at 148\,GHz found in the maps on the
celestial equator described in \cite{Hasselfield2012}. As an important
part of this process we measure the ``purity'' of the ACT SZ candidate
population over S82, that is, the fraction of real clusters as a
function of SZ detection significance.

\subsection{Cluster Confirmation Criteria}

\begin{figure}
\centerline{\includegraphics[width=3.9in]{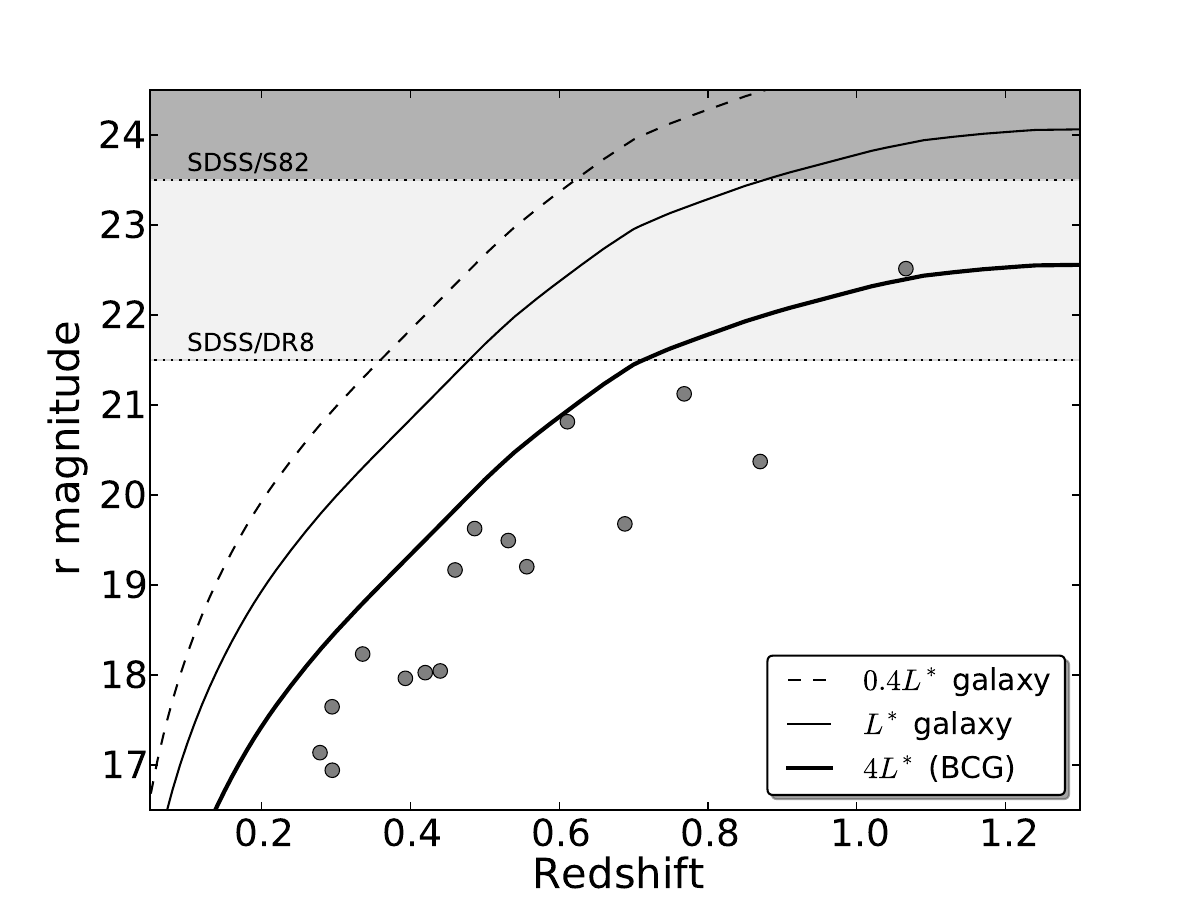}}
\caption{The observed $r$-band magnitudes of $L^*$, $0.4L^*$ and
  $4L^*$ (BCG) early-type galaxies as a function of redshift. We use
  $L^*$ as defined \cite{Blanton-03} for the population of red
  galaxies at $z=0.1$ and allow it to passively evolve with
  redshift. We show in gray the 50\% completeness limits for the
  SDSS/S82 and DR8 data for galaxies from \cite{Annis2011} reaching
  $r=23.5$ and $r=21.5$ respectively. For comparison we also show as
  gray circles the observed $r$-band magnitudes for the BCGs in the SZ
  southern sample from the imaging reported in \cite{Menanteau-SZ}. }
 \label{fig:mr_z}
\end{figure}
\vspace{0.5cm}

Our confirmation procedure builds upon our previous work on the ACT
southern sample \citep{Menanteau-SZ} and takes advantage of the
contiguous and deeper optical coverage available from S82, which allows
the systematic and rapid investigation of {\em all} SZ cluster
candidates, unlike for the 2008 ACT data and associated follow-up.
The procedure consists of searching for an optical cluster associated
with each candidate's SZ decrement. This is relatively
straightforward, since in concordance $\Lambda$CDM cosmology the halo
mass function \citep[e.g.,][]{Tinker2008} predicts that around 90\% of
massive clusters (i.e.,
$M_{200a}>3\times10^{14}\,h_{70}^{-1}M_\odot$), such as the ones that
make up the current generation of SZ samples, will lie below
$z\approx0.8$ and are therefore accessible for intermediate-depth
optical imaging such as in the S82 data set.

The optical confirmation requires the detection of a brightest cluster
galaxy (BCG) and an accompanying red sequence of cluster members,
which are typically early-type galaxies with luminosities less than
$L^*$ (the characteristic Schechter luminosity). 
In Section~\ref{sec:membership} we discuss our richness criterion for
optical confirmation of the sample.
We use the completeness limits estimated from simulations by
\cite{Annis2011} to determine how far in redshift we can ``see''
massive clusters in S82. For this, we compare the completeness limits
of S82 observations to the expected and observed apparent magnitudes
of galaxies {\em in} clusters as a function of redshift.  We estimated
the expected apparent galaxy $r$-band magnitude as a function of
redshift using $L^*$ as defined for the population of red galaxies by
\cite{Blanton-03} at $z=0.1$ and allowing passive evolution according
to a solar metallicity \citet{BC03} $\tau=1.0$~Gyr burst model formed
at $z_f=5$.  We show this relation in Figure~\ref{fig:mr_z} for a
range of luminosities ($0.4L^*$, $L^*$ and $4L^*$) aimed at
representing the cluster members from the faint ones to the BCG. We
also show as different gray levels the 50\% completeness level as
determined by the simulations for the S82 and DR8 samples
\citep{Annis2011}.
Figure~\ref{fig:mr_z} also shows, for comparison, the apparent $r$-band
magnitude of BCGs in the ACT southern cluster sample \cite{Menanteau-SZ}.

We conclude that we can comfortably detect cluster BCGs in S82 up to
$z>1$ and outside S82 to $z=0.7$.  A cluster red sequence will be
confidently detected to somewhat lower redshifts, $z\approx0.8$
(S82) and $z\approx0.5$ (outside S82), thus satisfying our criteria
for optical cluster confirmation.
In summary, we search for a BCG and associated red-sequence around
each SZ candidate; if this condition is satisfied, we estimate the
redshift and richness for the cluster. If the cluster richness
satisfies the minimum richness criteria (see
Section~\ref{sec:membership}) we list the candidate as a real cluster.

\subsection{Cluster Redshift Determination}
\label{sec:redshifts}

\begin{figure}
\centerline{\includegraphics[width=3.6in]{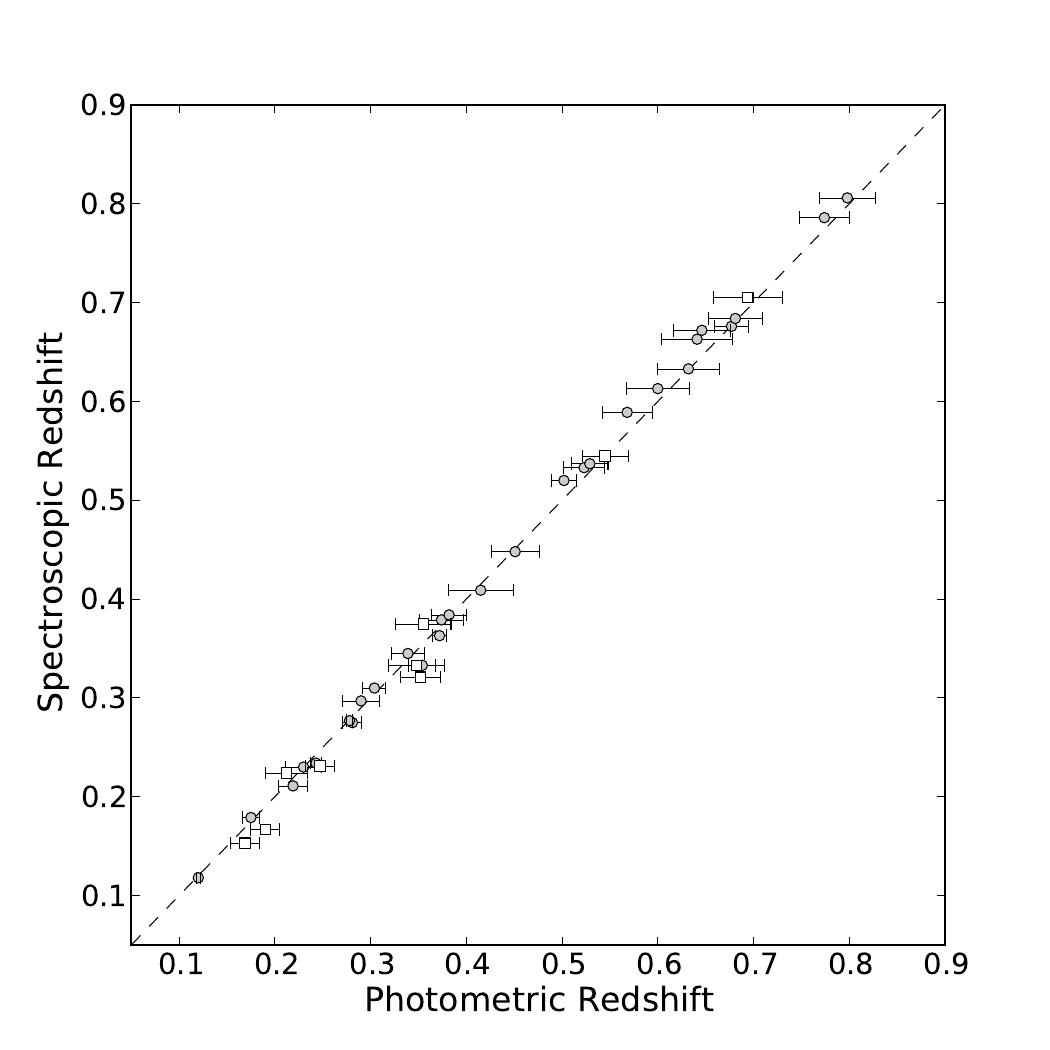}}  
\caption{Spectroscopic redshift versus photometric redshift for the
  sub-sample 37 of ACT equatorial clusters with known spectroscopic
  redshifts. Circles represent clusters from the S82 area while
  squares are system outside the S82 area. Error bars show the 68\% C.L.
  uncertainties on the cluster photometric redshift.}
 \label{fig:photo-z}
\end{figure}

In practice we perform the cluster confirmation by working on
$10'\times10'$ wide images centered on the position of the SZ
candidate that are created from stitching together nearby S82 tiles in
all 5 SDSS bands. Our inspection relies on a custom-created automated
software that enables us to interactively search for a BCG and its red
cluster sequence using our own implementation of the MaxBCG cluster
finder \citep{MaxBCG} algorithm, as described in \cite{SCSII} for the
Southern Cosmology Survey \citep[SCS;][]{SCSI}. 
Although our implementation of the MaxBCG cluster algorithm represents
our best effort to replicate the method as described in
\cite{MaxBCG}, the measured richness values should not be expected
to be exactly as in the original MaxBCG implementation due to slight
differences in the handling of photometric errors and background
subtraction.
This consists of
visually selecting the BCG and from that recorded position
iteratively choosing cluster member galaxies using the photometric
redshifts and a $3\sigma$ clipping algorithm within a local
self-defined color-magnitude relation.  For candidates with APO $K_S$
follow-up imaging we use 6 bands, which are limited to the
$\sim5'\times5'$ field-of-view (FOV) of NICFPS, but otherwise the
procedure is the same.
Our software aids the precise determination of the BCG by visually
flagging all early-type galaxies (i.e., galaxies SED types 0 and 1 from
BPZ) that are more luminous than $4L^*$ galaxies, with $L^*$ as defined above.
Once the BCG has been established, the next step in the optical
confirmation is to define the cluster redshift and color criteria to
be used in selecting cluster members as these are required to estimate
the richness of the cluster.

The determination of the cluster redshift is an iterative process,
using our custom-developed tools, that starts with the redshift of the
BCG as the initial guess for the cluster's redshift and center.
It then estimates the redshift as the mean value of the $N$ brightest
early-type galaxies (with $N=7$) within an inner radius of
$250\,h^{-1}$kpc (with $H_0=100{\rm\,km\,s^{-1}\,Mpc^{-1}} \,h$ as
defined by MaxBCG) and the redshift interval $\Delta z=0.045(1+z_c)$
where $z_c$ is the redshift of the cluster.
For redshift determination we use the $N$ brightest early-type galaxies,
rather than the BCG alone, to mitigate against biased photometric
redshifts resulting from BCGs with peculiar colors, such as in cool
core clusters.
The new redshift is used as input and the same procedure is repeated
until convergence on the redshift value is achieved, which usually
occurs in three iterations or less.
The selection of $N=7$ was informed by optimizing cluster redshifts for
systems with known spectroscopic redshifts. Uncertainties in the
cluster redshifts are determined via bootstrap resampling (10,000
times) of the galaxies selected for the redshift determination. We
also explored estimating errors using Monte Carlo realizations of
the sample which provided similar results.
We note that although our catalogs contain the spectroscopic redshifts
available from SDSS, in the procedure described here we only make use
of the photometric redshifts, in order to make a direct comparison with
the spectroscopic information.

Another important advantage of the overlap of ACT with S82 and SDSS is
that for all clusters at $z<0.3$ the BCG was spectroscopically
targeted by SDSS and has a spectroscopic redshift. Moreover, as
BCGs are very luminous objects, in several cases it was possible to
match them with a spectroscopic redshift from SDSS to
$z\approx0.5$. There are 25 ACT clusters in S82 for which a spectroscopic
redshift was available from SDSS for the BCG or the next brightest
galaxy in the cluster. For the ACT area outside S82, the CAS DR8
database provides imaging but no spectroscopic redshifts are available
from SDSS on this region.
Additionally, within the sample presented in this paper, 21 (18 are on
S82) S/N$>4.5$ SZ clusters have multi-object spectroscopic follow-up
observations using GMOS on Gemini-S as part of our program aimed at
obtaining dynamical masses for ACT clusters at $z>0.35$
\citep{Sifon2012}. The observations were carried out as part of our
ongoing programs (GS-2011B-C-1, GS-2012A-C-2 and GS-2012B-C-3) and
processed using our custom set of tools as described in
\cite{Sifon2012}. The full description of the ACT equatorial sample
follow-up with Gemini will be described in a future paper (Sif\'on et
al., in prep.). 
In Figure~\ref{fig:photo-z} we show that the photometric and spectroscopic
redshifts are in good agreement. Thus for clusters without
spectroscopic redshifts, up to $z\approx0.8$, we confirm that our
photometric ones will be quite accurate. For clusters at $z>0.9$, due
to the lack of spectroscopic redshifts, we can only assume that the SDSS
well-calibrated photometry provides robust estimates.
Both photometric and spectroscopic redshifts for the full cluster
sample are given in Tables~\ref{tab:S82sample} and
\ref{tab:DR8sample}.

\subsection{Defining Cluster Membership}
\label{sec:membership}

In order to have a richness measurement useful to compare across the
SZ cluster sample, one must define cluster membership. We follow a
similar procedure to that in \cite{SCSII}.
Once the redshift of the cluster is determined, we use BPZ-defined 
early-type galaxies within the same
$250\,h^{-1}$kpc radius and redshift interval $\Delta
z=0.045(1+z_c)$ as above to obtain a local self-defined
color-magnitude relation (CMR) for each color combination, $g-r$,
$r-i$, and $i-z$ ($z-K_s$ when available) for all cluster members,
using a $3\sigma$ clipping algorithm.
For the determination of cluster members we use the
spectroscopic redshift when available to define $z_c$.
We use these spatial and color criteria to determine $N_{\rm 1Mpc}$,
the number of galaxies within $1h^{-1}$\,Mpc of the cluster center as
defined by \cite{MaxBCG}. Formally, we compute $N_{\rm gal}=N_{\rm
  1Mpc}$ by including those galaxies within a projected $1h^{-1}$\,Mpc
from the cluster center and within $\Delta z=0.045(1+z_c)$ that
satisfy three conditions: (1) the galaxy must have the SED of an early
type according to BPZ, (2) it must have the appropriate color to be a
cluster member (i.e., colors within $3\sigma$ of the local CMR for all
color combinations) and (3) it must have the right luminosity, dimmer
than the BCG and brighter than $0.4L^*$.
Additionally we designate cluster members according to the estimated
cluster size $R_{200}$, defined as the radius at which the cluster
galaxy density is $200\Omega_m^{-1}$ times the mean space density of
galaxies in the present Universe. We estimated the scaled radius
$R_{200}$ using the empirical relation from \cite{Hansen-05},
$R_{200}= 0.156N_{\rm 1Mpc}^{0.6} h^{-1}$Mpc which is derived from the
SDSS. Hence $N_{200}$ is the number of galaxies satisfying the above
conditions within $R_{200}$.
We note however, that our ability to uniformly select cluster members
to $0.4L^*$ depends on the imaging depth of the data available. From
Figure~\ref{fig:mr_z} we infer that we can detect $0.4L^*$ galaxies
to $z\approx0.6$ and $z\approx0.4$ for clusters inside and outside of
S82 region respectively. Beyond this redshift range our richness
values underestimate the true values.
We caution the reader that beyond this redshift range our richness
values underestimate the true values as we do not attempt to correct
for the incompleteness of detecting galaxies expected above $0.4L^*$
at our limiting magnitude.

For our richness measurements we estimated the galaxy background
contamination and implemented an appropriate background subtraction
method following the same procedure described in \cite{SCSI} (see
section 3.1).
We use a statistical removal of unrelated field galaxies with similar
colors and redshifts that were projected along the line of sight to
each cluster. We estimate the surface number density of ellipticals in
an annulus surrounding the cluster (within $4<r<9\,h^{-1}$Mpc) with
the same $\Delta z$ as above and the same colors as the cluster
members. We measure this background contribution around the outskirts
of each cluster and obtain a corrected value $N_{\rm gal}$ which is
used to compute $R_{200}$ and then a corresponding value of
$N_{200}$. The magnitude of the correction ranges between $10\%-40$\%
depending on the cluster richness. For the clusters confirmed using
APO observations, the smaller FOV of NICFPS precludes us from making a
proper background correction for the $N_{\rm gal}$ estimate. Instead
we choose a conservative 40\% correction factor. In the few cases
where the cluster is located near the edge of the optical coverage of
S82 and the projected area of a $1h^{-1}$\,Mpc aperture is not fully
contained within the optical data we scale up $N_{\rm gal}$ by the
fraction of the missing area. We will refer to the corrected values
hereafter.

The measured richness value, $N_{\rm gal}$, was used in addition to
the presence of a BCG and accompanying red sequence to optically
confirm cluster candidates; we require a numerical value of $N_{\rm
  gal}>15$. In practice this additional constraint resulted in the
removal of only one candidate.
In Tables~\ref{tab:S82sample} and \ref{tab:DR8sample} we
present the $N_{\rm gal}$ estimated for the S82 and DR8 sample
respectively.

\begin{figure*}
\centerline{
\includegraphics[width=3.5in]{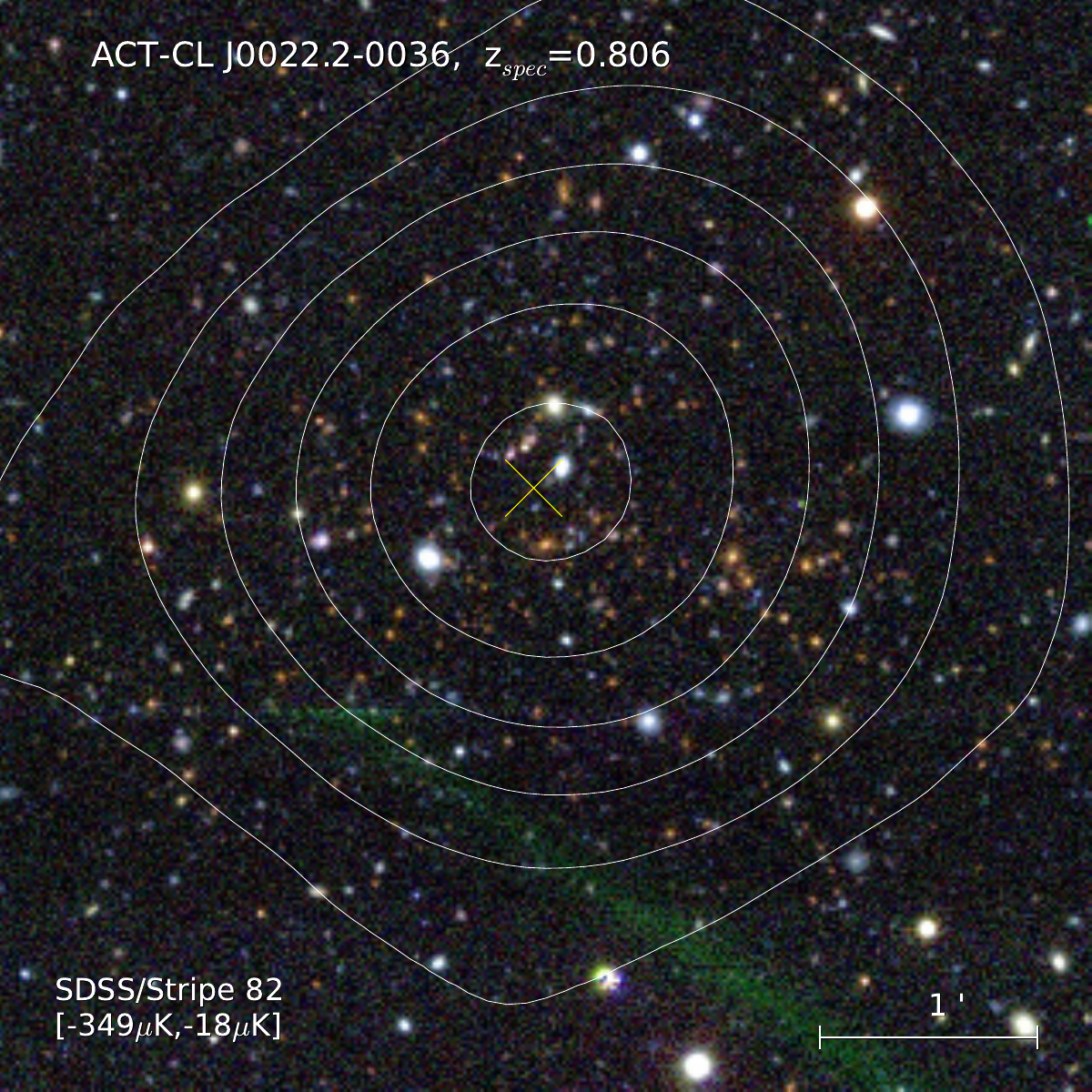}  
\includegraphics[width=3.5in]{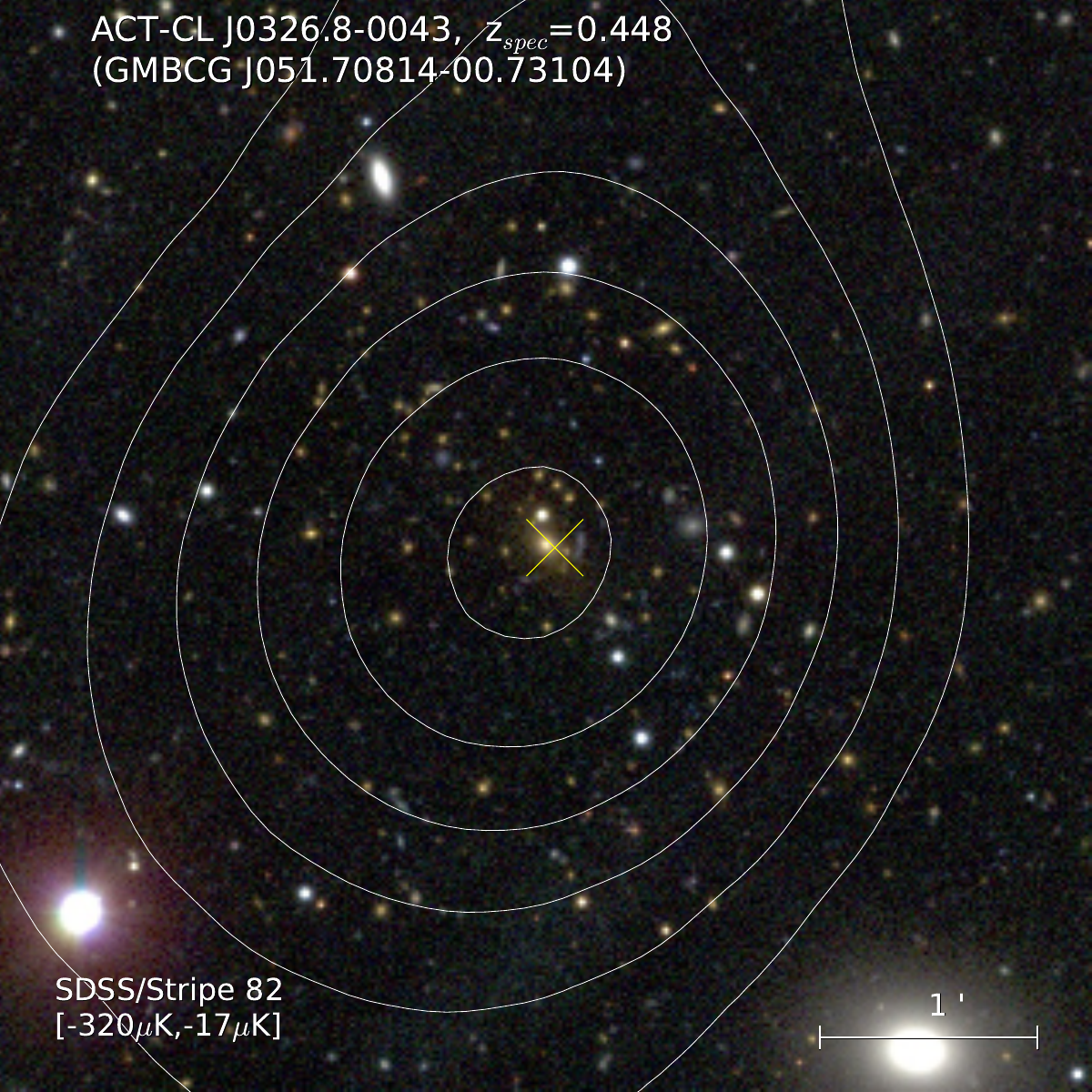}
}
\centerline{
\includegraphics[width=3.5in]{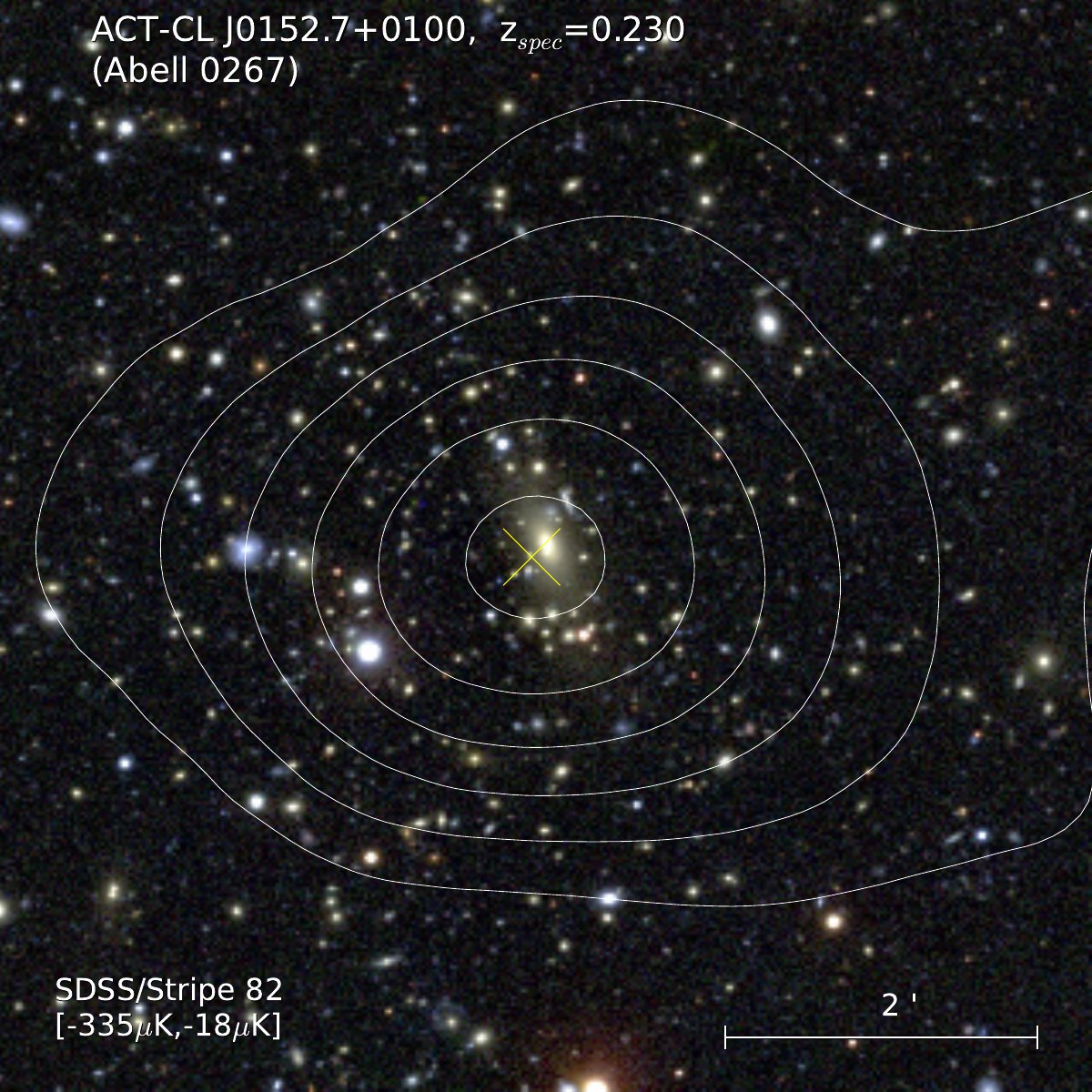}
\includegraphics[width=3.5in]{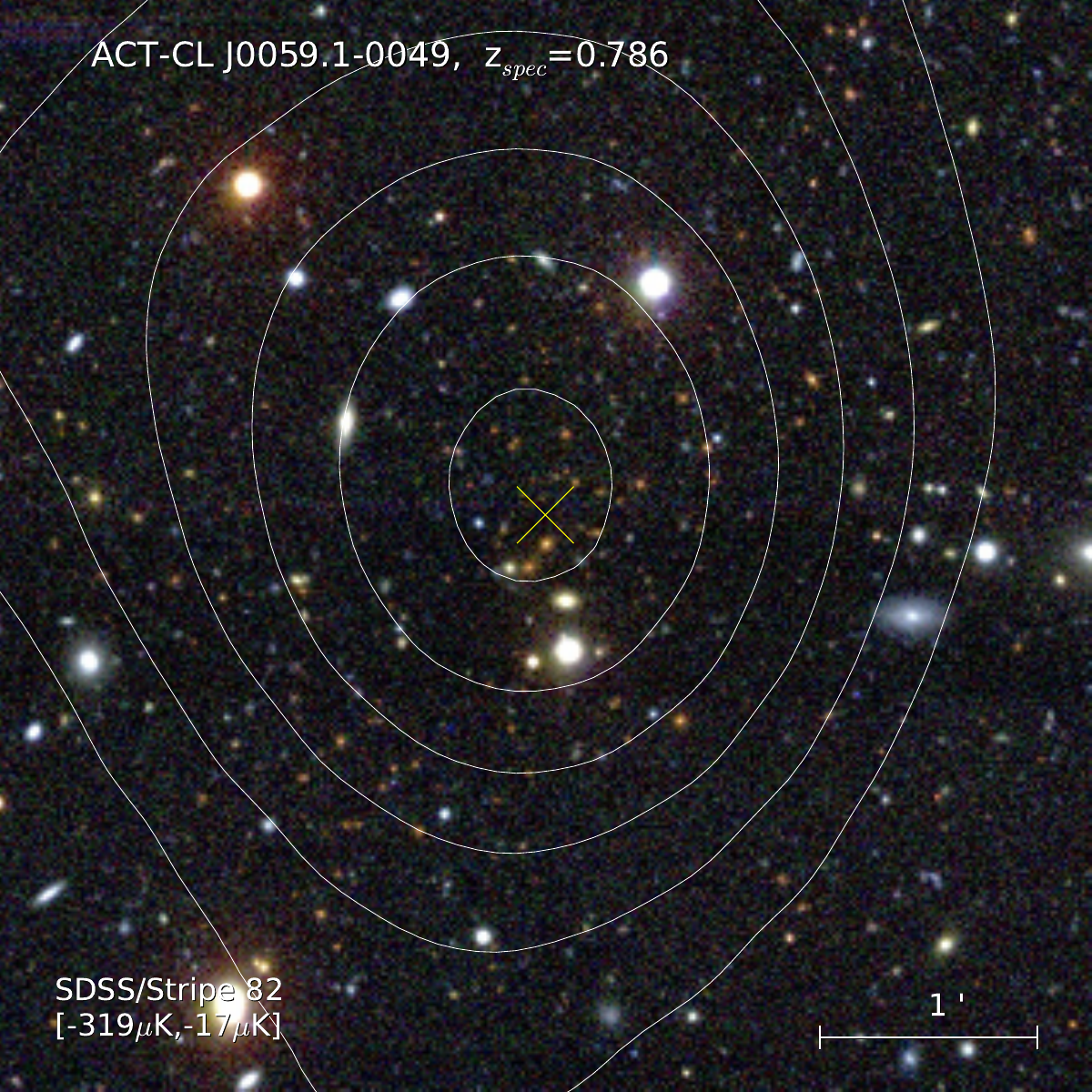}
}
\caption{Composite color images for 4 ACT SZ clusters optically
  confirmed using the S82 imaging. The horizontal bar shows the scale
  of the images, where north is up and east is left. White contours
  show the 148~GHz SZ maps with the minimum and maximum levels, in
  $\mu$K, displayed between brackets. The yellow cross shows the
  location of the centroid of the SZ detection.}
\label{fig:S82clustersa}
\end{figure*}

\begin{figure*}
\centerline{
\includegraphics[width=3.5in]{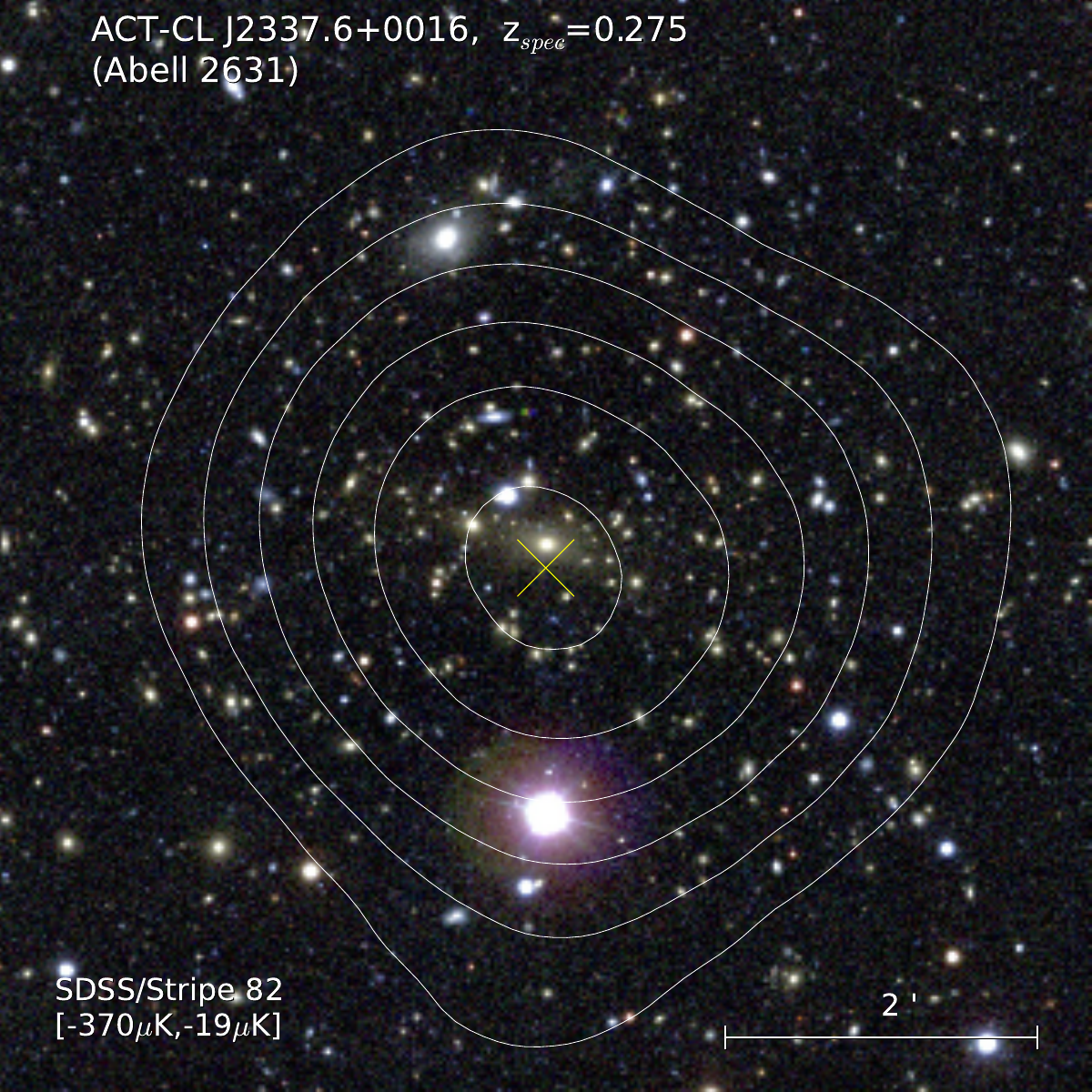}
\includegraphics[width=3.5in]{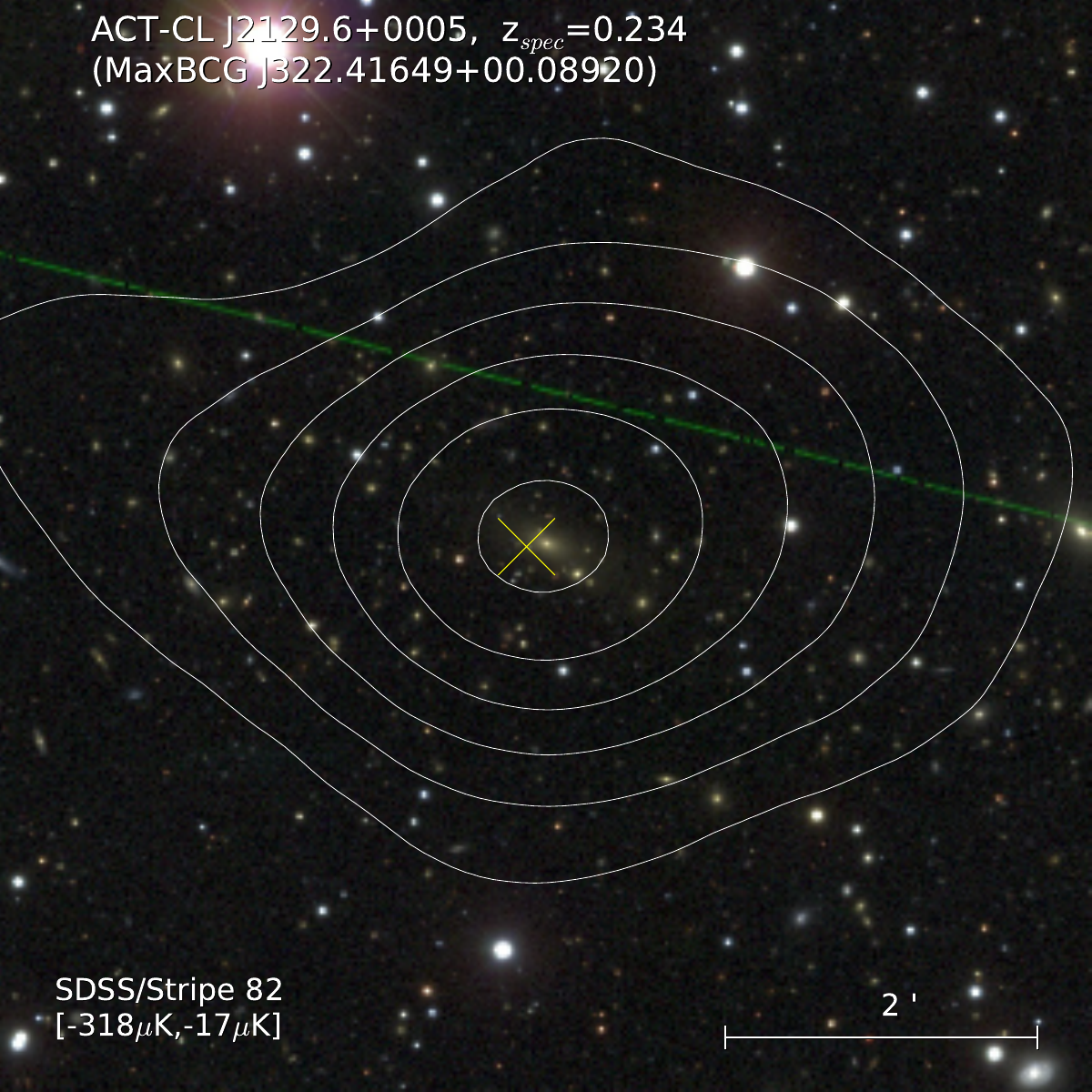}
}
\centerline{                  
\includegraphics[width=3.5in]{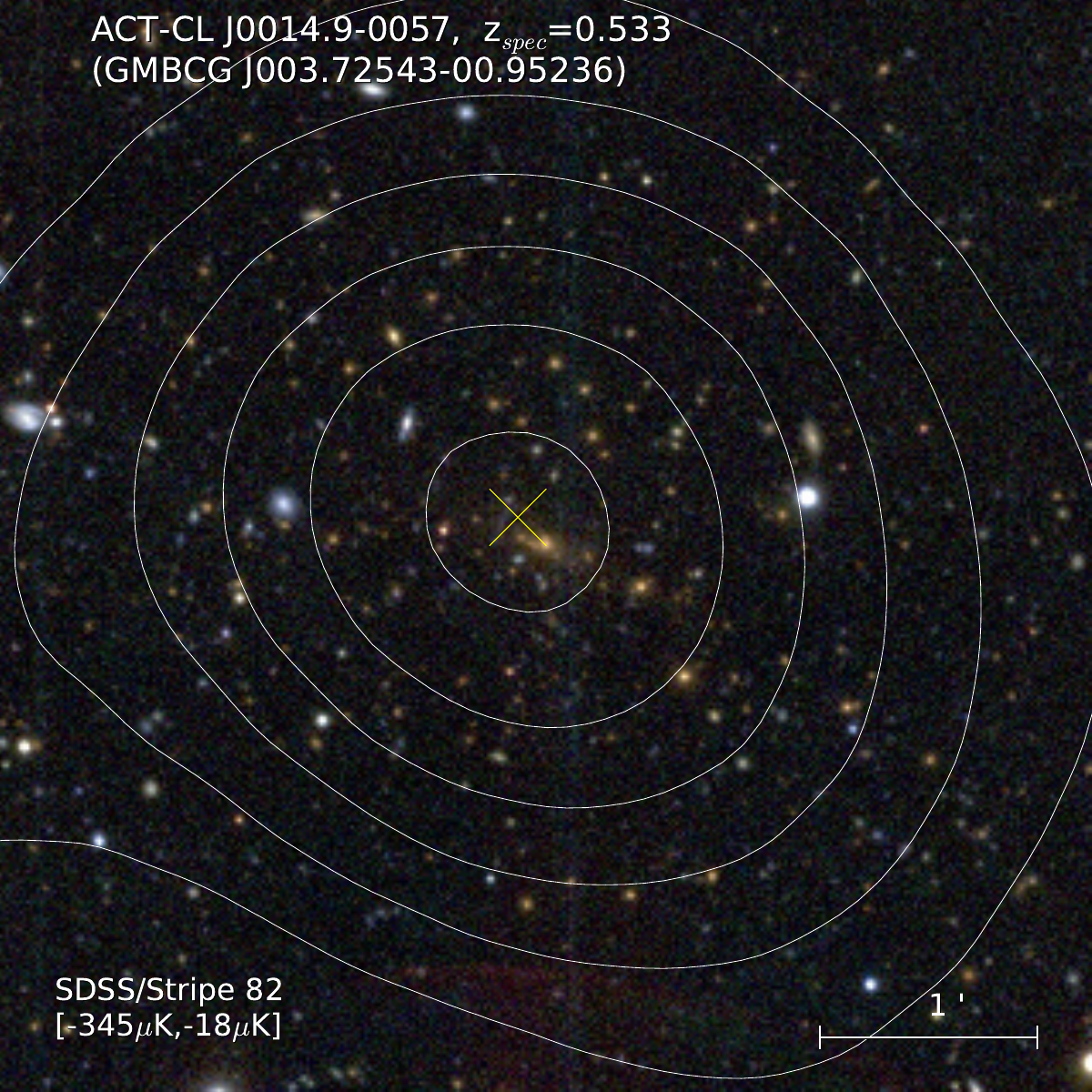}
\includegraphics[width=3.5in]{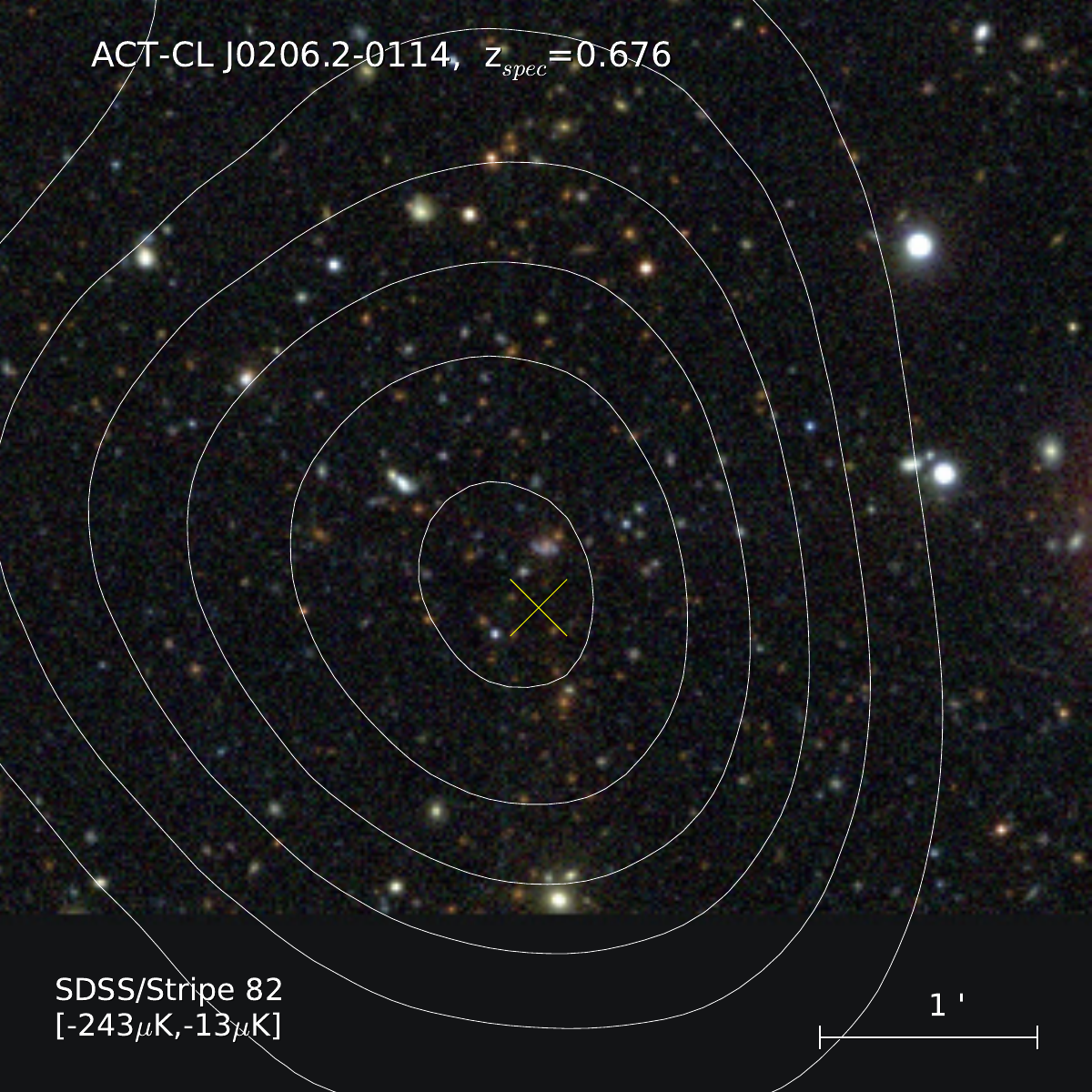}
}
\caption{Composite color images for 4 ACT SZ clusters optically
  confirmed using the S82 imaging. The horizontal bar shows the scale
  of the images, where north is up and east is left. White contours
  show the 148~GHz SZ maps with the minimum and maximum levels, in
  $\mu$K, displayed between brackets. The yellow cross shows the
  location of the centroid of the SZ detection.}
\label{fig:S82clustersb}
\end{figure*}

\section{The ACT Equatorial SZ Cluster Sample}

Our optical confirmation of SZ candidates has resulted in a new sample
of 68 clusters: 49 systems are located in the area
overlapping with S82 and 19 clusters on the area that overlaps with
the shallower DR8 data.

\renewcommand{\arraystretch}{1.05} 
\begin{deluxetable*}{lcccclccl} 
\tablecaption{Optically confirmed ACT Equatorial clusters on Stripe 82} 
\tablehead{
\colhead{ACT Descriptor} & 
\colhead{R.A. (J2000)} & 
\colhead{Dec. (J2000)} &
\colhead{$z$-spec} &
\colhead{$z$-photo} &
\colhead{$N_{\rm gal}$} &
\colhead{SNR} &
\colhead{BCG distance} &
\colhead{Alternative Name} \\
\colhead{} & 
\colhead{} & 
\colhead{} & 
\colhead{} & 
\colhead{} & 
\colhead{($1h^{-1}$\,Mpc)} & 
\colhead{(148\,GHz)} & 
\colhead{(Mpc $h_{70}^{-1}$)} & 
\colhead{} 
}
\startdata
ACT-CL~J0022.2$-$0036 &  00:22:13.0 & $-$00:36:33.8 &   0.805 \tablenotemark{$\dagger$}  &  $0.80 \pm 0.03$ & $65.9\pm8.1^\ast$ & 9.8 & 0.124 &  \\
ACT-CL~J0326.8$-$0043 &  03:26:49.9 & $-$00:43:51.7 &   0.448 \tablenotemark{$\ddagger$} &  $0.45 \pm 0.03$ & $41.7\pm6.5$ & 9.1 & 0.014 & GMBCG J051.70814-00.73104 \tablenotemark{a} \\
ACT-CL~J0152.7$+$0100 &  01:52:41.9 & $+$01:00:25.5 &   0.230 \tablenotemark{$\star$}    &  $0.23 \pm 0.02$ & $67.2\pm8.2$ & 9.0 & 0.026 & Abell 0267 \tablenotemark{b} \\
ACT-CL~J0059.1$-$0049 &  00:59:08.5 & $-$00:50:05.7 &   0.786 \tablenotemark{$\dagger$}  &  $0.77 \pm 0.03$ & $41.0\pm6.4^\ast$ & 8.4 & 0.064 &  \\
ACT-CL~J2337.6$+$0016 &  23:37:39.7 & $+$00:16:16.9 &   0.275 \tablenotemark{$\star$}    &  $0.28 \pm 0.01$ & $57.8\pm7.6$ & 8.2 & 0.036 & Abell 2631 \tablenotemark{b} \\
ACT-CL~J2129.6$+$0005 &  21:29:39.9 & $+$00:05:21.1 &   0.234 \tablenotemark{$\star$}    &  $0.24 \pm 0.01$ & $35.0\pm5.9$ & 8.0 & 0.028 & RX J2129.6+0005 \tablenotemark{c} \\
ACT-CL~J0014.9$-$0057 &  00:14:54.1 & $-$00:57:08.4 &   0.533 \tablenotemark{$\ddagger$} &  $0.52 \pm 0.02$ & $56.2\pm7.5$ & 7.8 & 0.070 & GMBCG J003.72543-00.95236 \tablenotemark{d} \\
ACT-CL~J0206.2$-$0114 &  02:06:13.4 & $-$01:14:17.0 &   0.676 \tablenotemark{$\dagger$}  &  $0.68 \pm 0.02$ & $68.4\pm9.5^\ast$ & 6.9 & 0.123 &  \\
ACT-CL~J0342.0$+$0105 &  03:42:02.1 & $+$01:05:07.5 & \nodata                            &  $1.07 \pm 0.06$ & $41.2\pm6.3^\ast$ & 5.9 & 0.248 &  \\
ACT-CL~J2154.5$-$0049 &  21:54:32.3 & $-$00:49:00.4 &   0.488 \tablenotemark{$\dagger$}  &  $0.48 \pm 0.02$ & $56.9\pm7.5$ & 5.9 & 0.090 & WHL J215432.2-004905 \tablenotemark{e} \\
ACT-CL~J0218.2$-$0041 &  02:18:16.8 & $-$00:41:41.8 &   0.672 \tablenotemark{$\dagger$}  &  $0.65 \pm 0.03$ & $39.2\pm6.3^\ast$ & 5.8 & 0.262 &  \\
ACT-CL~J0223.1$-$0056 &  02:23:10.0 & $-$00:57:08.9 &   0.663 \tablenotemark{$\ddagger$} &  $0.64 \pm 0.04$ & $50.5\pm7.1^\ast$ & 5.8 & 0.159 & in GMB2011 \\
ACT-CL~J2050.5$-$0055 &  20:50:29.7 & $-$00:55:40.6 &   0.622 \tablenotemark{$\ddagger$} &  $0.60 \pm 0.03$ & $38.6\pm6.2^\ast$ & 5.6 & 0.098 & in GMB2011 \\
ACT-CL~J0044.4$+$0113 &  00:44:25.6 & $+$01:12:48.7 & \nodata                            &  $1.11 \pm 0.03$ & $73.0\pm8.5^\ast$ & 5.5 & 0.258 &  \\
ACT-CL~J0215.4$+$0030 &  02:15:28.5 & $+$00:30:37.3 &   0.865 \tablenotemark{$\dagger$}  &  $0.73 \pm 0.03$ & $29.5\pm3.8^\ast$ & 5.5 & 0.046 &  \\
ACT-CL~J0256.5$+$0006 &  02:56:33.7 & $+$00:06:28.8 &   0.363 \tablenotemark{$\ddagger$} &  $0.37 \pm 0.01$ & $39.8\pm6.3$ & 5.4 & 0.113 & RX J0256.5+0006 \tablenotemark{c} \\
ACT-CL~J0012.0$-$0046 &  00:12:01.8 & $-$00:46:34.5 & \nodata                            &  $1.36 \pm 0.06$ & $29.2\pm5.3^\ast$ & 5.3 & 0.313 &  \\
ACT-CL~J0241.2$-$0018 &  02:41:15.4 & $-$00:18:41.0 &   0.684 \tablenotemark{$\star$}    &  $0.68 \pm 0.03$ & $50.5\pm7.1^\ast$ & 5.1 & 0.040 &  \\
\\                                                                                                                         
\hline                                                                                                                     
\\                                                                                                                         
ACT-CL~J0127.2$+$0020 &  01:27:16.6 & $+$00:20:40.9 &   0.379 \tablenotemark{$\ddagger$} &  $0.37 \pm 0.02$ & $64.8\pm8.1$ & 5.1 & 0.075 & GMBCG J021.81939+00.34469 \tablenotemark{a} \\
ACT-CL~J0348.6$+$0029 &  03:48:36.7 & $+$00:29:33.0 &   0.297 \tablenotemark{$\star$}    &  $0.29 \pm 0.02$ & $29.4\pm5.4$ & 5.0 & 0.142 & GMBCG J057.17821+00.48718 \tablenotemark{a} \\
ACT-CL~J0119.9$+$0055 &  01:19:58.1 & $+$00:55:33.6 & \nodata                            &  $0.72 \pm 0.03$ & $21.5\pm3.3^\ast$ & 5.0 & 0.218 &  \\
ACT-CL~J0058.0$+$0030 &  00:58:05.7 & $+$00:30:58.1 & \nodata                            &  $0.76 \pm 0.02$ & $47.0\pm6.8^\ast$ & 5.0 & 0.199 &  \\
ACT-CL~J0320.4$+$0032 &  03:20:29.7 & $+$00:31:53.7 &   0.384 \tablenotemark{$\star$}    &  $0.38 \pm 0.02$ & $55.9\pm7.5$ & 4.9 & 0.158 & GMBCG J050.12410+00.53157 \tablenotemark{a} \\
ACT-CL~J2302.5$+$0002 &  23:02:35.0 & $+$00:02:34.2 &   0.520 \tablenotemark{$\ddagger$} &  $0.50 \pm 0.01$ & $61.4\pm7.8$ & 4.9 & 0.080 & GMBCG J345.64608+00.04281 \tablenotemark{a} \\
ACT-CL~J2055.4$+$0105 &  20:55:23.2 & $+$01:06:07.5 &   0.408 \tablenotemark{$\ddagger$} &  $0.41 \pm 0.03$ & $37.7\pm6.1$ & 4.9 & 0.233 & GMBCG J313.84687+01.10212 \tablenotemark{a} \\
ACT-CL~J0308.1$+$0103 &  03:08:12.1 & $+$01:03:15.0 &   0.633 \tablenotemark{$\star$}    &  $0.63 \pm 0.03$ & $41.1\pm6.4^\ast$ & 4.8 & 0.174 &  \\
ACT-CL~J0336.9$-$0110 &  03:36:57.1 & $-$01:09:48.3 & \nodata                            &  $1.32 \pm 0.05$ & $29.1\pm5.1^\ast$ & 4.8 & 0.277 &  \\
ACT-CL~J0219.8$+$0022 &  02:19:50.4 & $+$00:22:14.9 &   0.537 \tablenotemark{$\star$}    &  $0.53 \pm 0.02$ & $59.0\pm7.7$ & 4.7 & 0.191 & GMBCG J034.95781+00.37385 \tablenotemark{a} \\
ACT-CL~J0348.6$-$0028 &  03:48:39.5 & $-$00:28:16.9 &   0.345 \tablenotemark{$\star$}    &  $0.34 \pm 0.02$ & $56.9\pm7.5$ & 4.7 & 0.095 & GMBCG J057.14850-00.43348 \tablenotemark{a} \\
ACT-CL~J2351.7$+$0009 &  23:51:44.6 & $+$00:09:16.2 & \nodata                            &  $0.99 \pm 0.03$ & $76.0\pm8.7^\ast$ & 4.7 & 0.039 &  \\
ACT-CL~J0342.7$-$0017 &  03:42:42.6 & $-$00:17:08.3 &   0.310 \tablenotemark{$\star$}    &  $0.30 \pm 0.01$ & $36.3\pm6.0$ & 4.6 & 0.132 & GMBCG J055.67773-00.28564 \tablenotemark{a} \\
ACT-CL~J0250.1$+$0008 &  02:50:08.4 & $+$00:08:16.4 & \nodata                            &  $0.78 \pm 0.03$ & $32.7\pm5.7^\ast$ & 4.5 & 0.084 &  \\
ACT-CL~J2152.9$-$0114 &  21:52:55.6 & $-$01:14:53.2 & \nodata                            &  $0.69 \pm 0.02$ & $22.7\pm3.9^\ast$ & 4.4 & 0.156 &  \\
ACT-CL~J2130.1$+$0045 &  21:30:08.8 & $+$00:46:48.3 & \nodata                            &  $0.71 \pm 0.04$ & $21.5\pm3.3^\ast$ & 4.4 & 0.554 &  \\
ACT-CL~J0018.2$-$0022 &  00:18:18.4 & $-$00:22:45.8 & \nodata                            &  $0.75 \pm 0.04$ & $27.8\pm5.3^\ast$ & 4.4 & 0.393 &  \\
ACT-CL~J0104.8$+$0002 &  01:04:55.3 & $+$00:03:36.2 &   0.277 \tablenotemark{$\star$}    &  $0.28 \pm 0.00$ & $64.4\pm8.0$ & 4.3 & 0.235 & MaxBCG J016.23069+00.06007 \tablenotemark{d} \\
ACT-CL~J0017.6$-$0051 &  00:17:37.6 & $-$00:52:42.0 &   0.211 \tablenotemark{$\star$}    &  $0.22 \pm 0.01$ & $38.3\pm6.2$ & 4.2 & 0.268 & MaxBCG J004.40671-00.87833 \tablenotemark{d} \\
ACT-CL~J0230.9$-$0024 &  02:30:53.8 & $-$00:24:40.9 & \nodata                            &  $0.44 \pm 0.03$ & $19.9\pm4.5$ & 4.2 & 0.158 & WHL J023055.3-002549 \tablenotemark{e} \\
ACT-CL~J0301.1$-$0110 &  03:01:12.0 & $-$01:10:47.7 & \nodata                            &  $0.53 \pm 0.04$ & $24.5\pm5.0$ & 4.2 & 0.260 & in GMB2011 \\
ACT-CL~J0051.1$+$0055 &  00:51:12.8 & $+$00:55:54.4 & \nodata                            &  $0.69 \pm 0.03$ & $20.5\pm3.2^\ast$ & 4.2 & 0.417 &  \\
ACT-CL~J0245.8$-$0042 &  02:45:51.7 & $-$00:42:16.4 &   0.179 \tablenotemark{$\star$}    &  $0.17 \pm 0.01$ & $40.2\pm6.3$ & 4.1 & 0.038 & Abell 0381 \tablenotemark{b} \\
ACT-CL~J2051.1$+$0056 &  20:51:11.0 & $+$00:56:46.1 &   0.333 \tablenotemark{$\star$}    &  $0.35 \pm 0.01$ & $20.2\pm4.5$ & 4.1 & 0.066 & GMBCG J312.79620+00.94615 \tablenotemark{a} \\
ACT-CL~J2135.1$-$0102 &  21:35:12.0 & $-$01:03:00.1 & \nodata                            &  $0.33 \pm 0.01$ & $68.0\pm8.2$ & 4.1 & 0.242 & GMBCG J323.80039-01.04962 \tablenotemark{a} \\
ACT-CL~J0228.5$+$0030 &  02:28:30.4 & $+$00:30:35.7 & \nodata                            &  $0.72 \pm 0.02$ & $31.5\pm4.0^\ast$ & 4.0 & 0.182 &  \\
ACT-CL~J2229.2$-$0004 &  22:29:07.5 & $-$00:04:11.0 & \nodata                            &  $0.61 \pm 0.05$ & $16.4\pm3.9^\ast$ & 4.0 & 0.569 &  \\
ACT-CL~J2135.7$+$0009 &  21:35:39.5 & $+$00:09:57.1 &   0.118 \tablenotemark{$\star$}    &  $0.12 \pm 0.00$ & $75.3\pm8.7$ & 4.0 & 0.144 & Abell 2356 \tablenotemark{b} \\
ACT-CL~J2253.3$-$0031 &  22:53:24.2 & $-$00:30:30.8 & \nodata                            &  $0.54 \pm 0.01$ & $23.0\pm3.4$ & 4.0 & 0.488 &  \\
ACT-CL~J2220.7$-$0042 &  22:20:47.0 & $-$00:41:54.4 & \nodata                            &  $0.57 \pm 0.03$ & $34.5\pm5.9$ & 4.0 & 0.277 & in GMB2011 \\
ACT-CL~J0221.5$-$0012 &  02:21:36.6 & $-$00:12:19.8 &   0.589 \tablenotemark{$\star$}    &  $0.57 \pm 0.03$ & $21.2\pm4.6$ & 4.0 & 0.246 & in GMB2011 \\
\enddata
\label{tab:S82sample}
\tablecomments{R.A.\ and Dec.\ positions denote the BCG location in
  the optical images of the cluster from our confirmation
  procedure. . The SZ position was used to construct the ACT
  descriptor identifiers. Spectroscopic redshifts are reported when
  available and come from the DR8 spectroscopic database and our own
  follow-up with GMOS on Gemini South. The horizontal line denotes the
  demarcation for the SZ cluster sample with 100\% purity. Values of
  S/N are from \cite{Hasselfield2012}.}

\tablenotetext{$\dagger$}{Spectroscopic redshift from GMOS/Gemini (Sif\'on et al.,in prep)}
\tablenotetext{$\ddagger$}{Spectroscopic redshift from GMOS/Gemini and SDSS}
\tablenotetext{$\star$}{Spectroscopic redshift from SDSS}
\tablenotetext{a}{from \cite{GMBCG}}
\tablenotetext{b}{from \cite{Abell1958}}
\tablenotetext{c}{from \cite{RX}}
\tablenotetext{d}{from \cite{MaxBCG} }
\tablenotetext{e}{from \cite{WHL}}
\tablenotetext{$\ast$}{Denotes clusters at $z>0.6$ for which the $0.4L^*$ limit was not reached and hence richness values are underestimated.}
\end{deluxetable*}

\subsection{Clusters in Stripe 82}
\label{sec:S82}

\begin{figure*}
\centerline{
\includegraphics[width=3.5in]{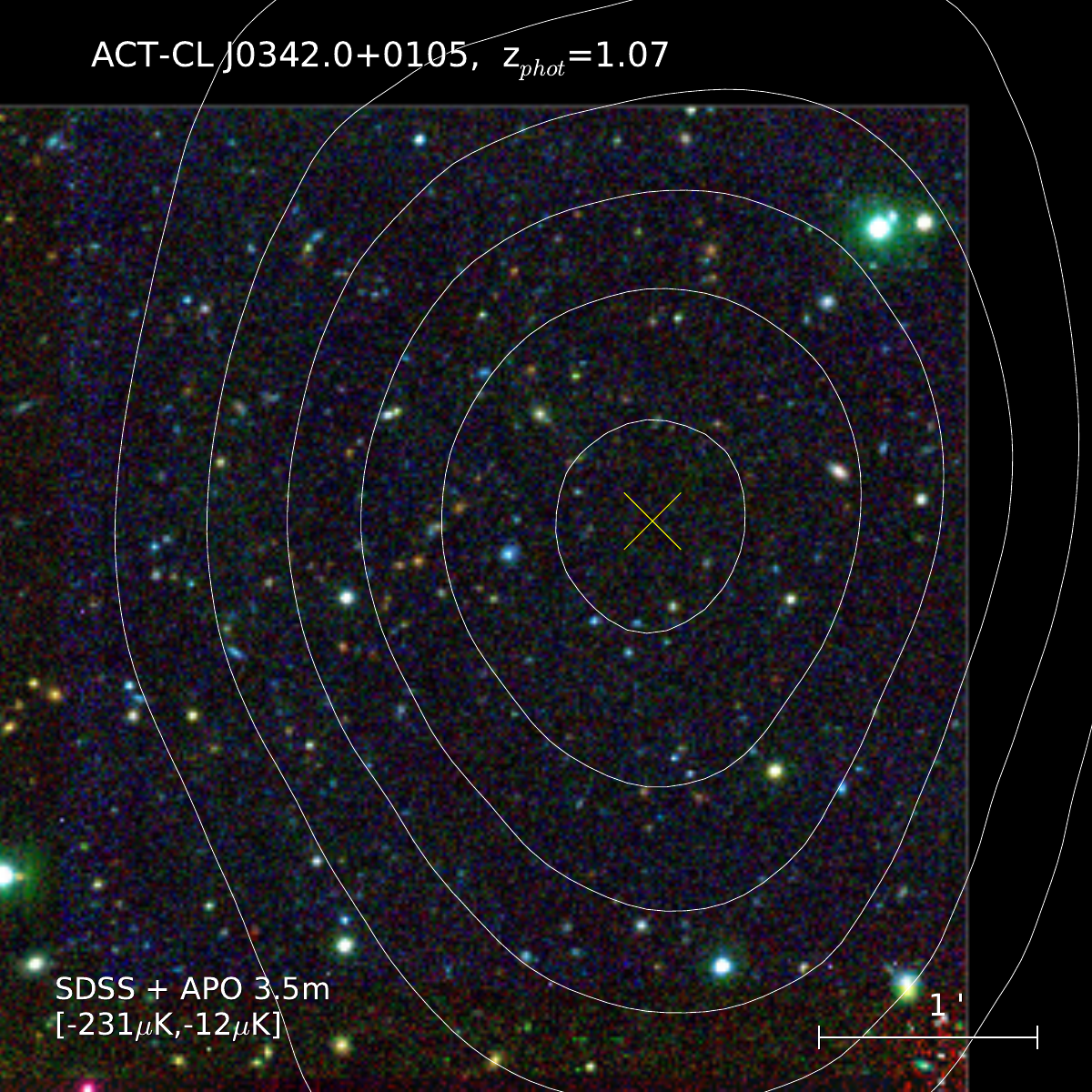}  
\includegraphics[width=3.5in]{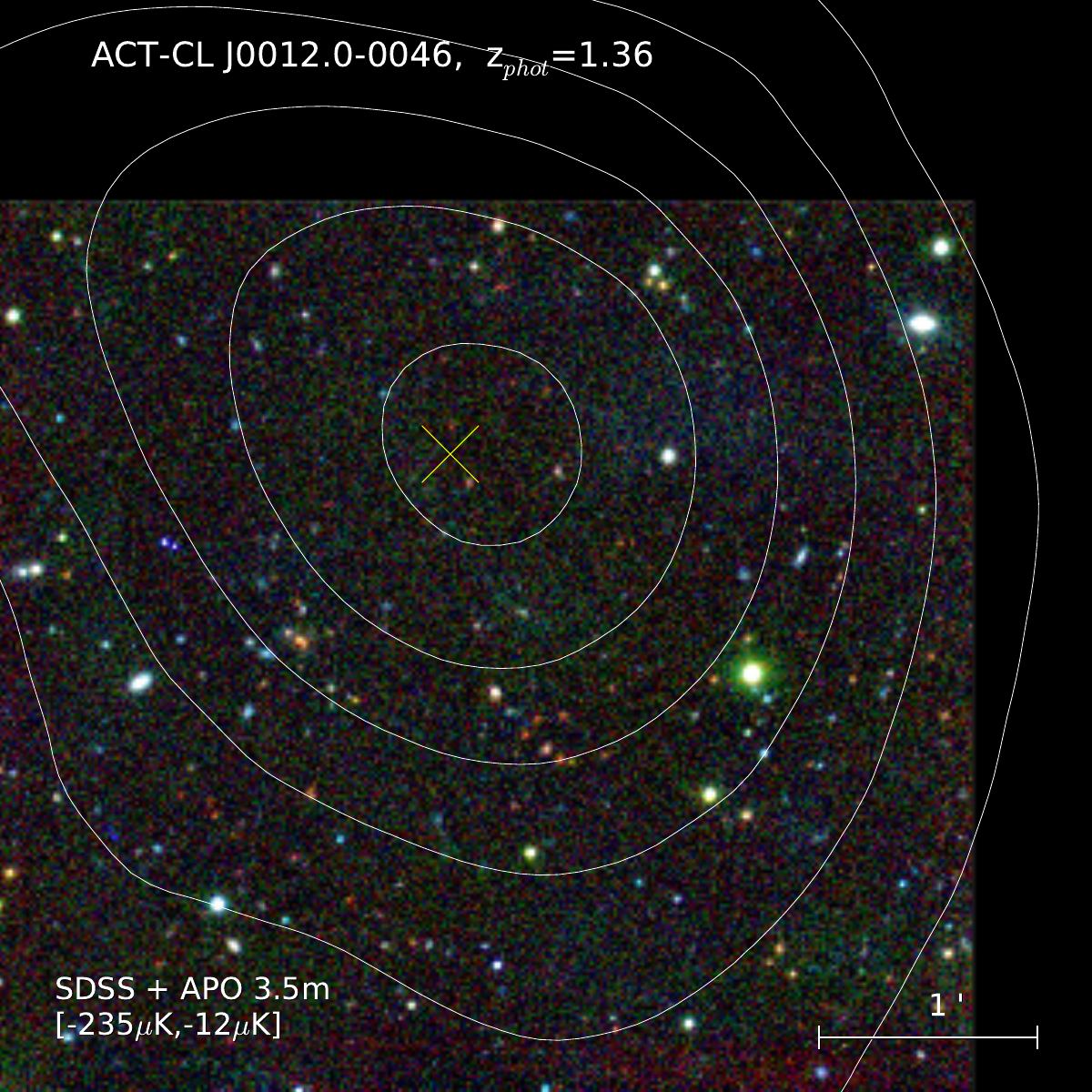}
}
\centerline{
\includegraphics[width=3.5in]{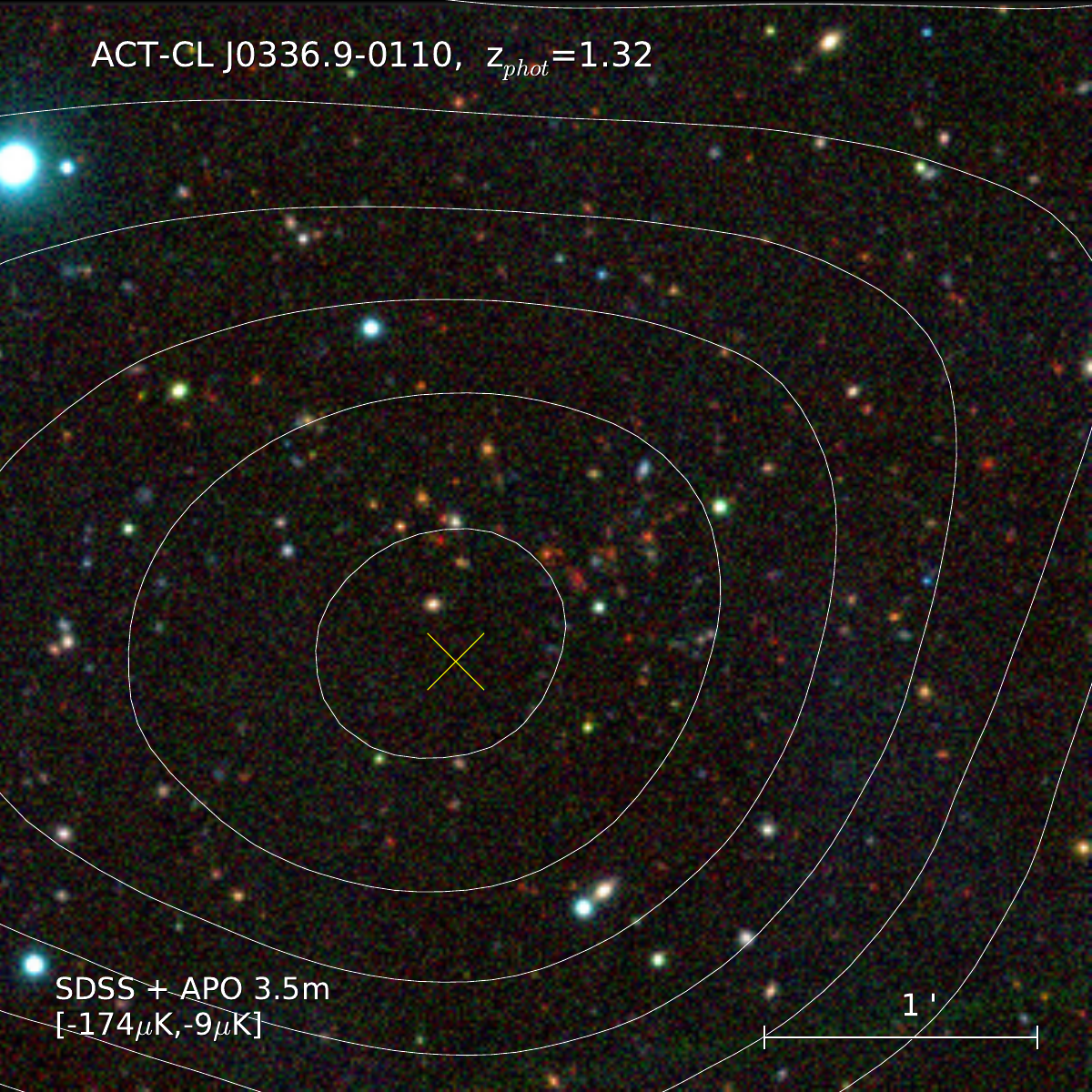}
\includegraphics[width=3.5in]{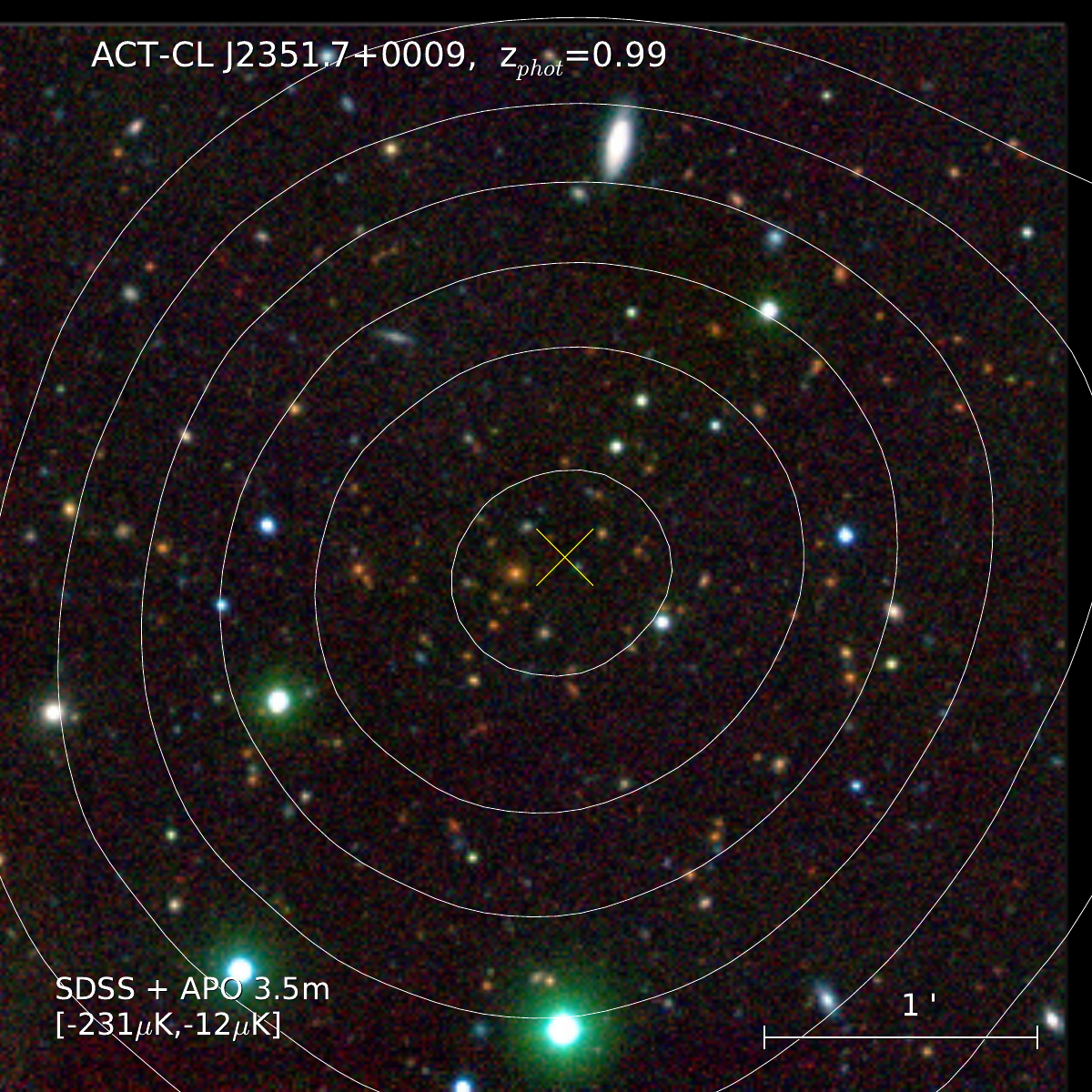}
}
\caption{Composite color images for 4 of the 5 high redshift ACT
  SZ clusters confirmed using optical (S82) and near-infrared (APO)
  imaging. The horizontal bar shows the scale of the images, where
  north is up and east is left. White contours show the 148~GHz SZ
  maps with the minimum and maximum levels, in $\mu$K, displayed
  between brackets. The yellow cross shows the location of the
  centroid of the SZ detection.}
\label{fig:APOclusters}
\end{figure*}

In Table~\ref{tab:S82sample} we present the 49 clusters in the
270\,deg$^2$ area in S82 along with their redshift information, BCG
positions and optical richness. In Figures~\ref{fig:S82clustersa}
and \ref{fig:S82clustersb} we show 8 examples of $z<1$ clusters
confirmed using the S82 imaging alone, while in
Figure~\ref{fig:APOclusters} we show examples of clusters confirmed
using the additional $K_S$-band APO imaging. Optical and NIR images
for the full sample are available at
\url{http://peumo.rutgers.edu/act/S82}.
We used NED\footnote{http://ned.ipac.caltech.edu} to search for
cluster counterparts for our sample using a $500\,h_{70}^{-1}$kpc
matching radius and found that a number of them are well-known $z<0.35$
clusters reported as part of the Abell \citep{Abell1958}, ROSAT
All-Sky Galaxy Cluster Survey \citep[NORAS;][]{RX} and MaxBCG
\citep{MaxBCG} catalogs. Also using NED we found matches for $z<0.55$
systems in the GMBCG \citep{GMBCG} and WHL \citep{WHL} optical cluster
catalogs. The GMBCG catalog is an improved version of the MaxBCG
method which used the SDSS DR7. In Table~\ref{tab:S82sample} we
designate the first reported alternative name for each system.
For higher redshift systems we compared our sample with the catalog
from \citet[][GMB2011]{Geach2011} which uses a cluster red sequence
algorithm on the same deep co-added S82 data used in this analysis to
detect clusters. We searched for counterparts using the same match
radius and found 5 previously reported GMB2011 systems at
$0.50<z<0.65$.
Beyond $z>0.65$ all SZ confirmed cluster in S82 represent new
discoveries, highlighting the power of the SZ effect to discover
massive galaxy clusters at high redshift. 
In summary, of the 49 ACT SZ-selected clusters from S82, 22 are new
and lie at $z>0.54$.

Our APO follow-up campaign aided in the confirmation of five new
clusters at $z\geq1$ over the S82 region by the addition of the $K_S$
imaging, described in Section~\ref{sec:APO}, to the 5 optical bands.
In Figures~\ref{fig:APOclusters} and \ref{fig:J0044} we show the
optical and NIR composite images of the 5 clusters at $z\geq1$.
In Table~\ref{tab:APO} we present a summary for the 5 new
clusters confirmed with the help of the NIR imaging.

\subsection{Additional Clusters Outside Stripe 82}

In Table~\ref{tab:DR8sample} we present the sample of 19 optically
confirmed clusters using the $ugriz$ imaging from the SDSS DR8 where
we provide the same information as for the S82 sample above.  The
shallower coverage over the area beyond S82 only allows us to present
optical confirmations for an incomplete subsample. As we see from
Figure~\ref{fig:mr_z}, the imaging depth of the DR8 data set can only
``see'' $L^*$ galaxies up to $z\approx0.5$. Moreover, the DR8
footprint does not fully cover the ACT equatorial region.
Within this sky region, which contains 10 new clusters, is located the
most significant SZ detection of the whole ACT equatorial sample,
\J2327 which we discuss in detail in Section~\ref{sec:J2327}.

An approved dedicated optical and NIR follow-up program using the SOAR
4.1-m and APO 3.5-m telescopes in 2012B will provide a more uniform
and complete cluster sample for the remaining area outside S82.

\renewcommand{\arraystretch}{1.05} 
\begin{deluxetable*}{lcclclccl} 
\tablecaption{Optically confirmed ACT Equatorial clusters outside
  Stripe 82, with DR8 coverage} 
\tablehead{
\colhead{ACT Descriptor} & 
\colhead{R.A. (J2000)} & 
\colhead{Dec. (J2000)} &
\colhead{$z$-spec} &
\colhead{$z$-photo} &
\colhead{$N_{\rm gal}$} &
\colhead{SNR} &
\colhead{BCG distance} &
\colhead{Alternative Name} \\
\colhead{} & 
\colhead{} & 
\colhead{} & 
\colhead{} & 
\colhead{} & 
\colhead{($1h^{-1}$\,Mpc)} & 
\colhead{(148\,GHz)} & 
\colhead{(Mpc $h_{70}^{-1}$)} & 
\colhead{} 
}
\startdata
ACT-CL~J2327.4$-$0204 &  23:27:27.6 & $-$02:04:37.4 &   0.705                   &  $0.69 \pm 0.04$ & $ 61.7\pm 7.9^\ast$ & 13.1 & 0.028 & RCS2 2327 \tablenotemark{1} \\
ACT-CL~J2135.2$+$0125 &  21:35:18.7 & $+$01:25:27.0 &   0.231 \tablenotemark{3} &  $0.25 \pm 0.01$ & $ 57.6\pm 7.6$ &  9.3 & 0.184 & Abell 2355 \tablenotemark{2} \\
ACT-CL~J0239.8$-$0134 &  02:39:53.1 & $-$01:34:56.0 &   0.375 \tablenotemark{4} &  $0.35 \pm 0.03$ & $ 84.0\pm 9.2$ &  8.8 & 0.121 & Abell 0370 \tablenotemark{2} \\
ACT-CL~J2058.8$+$0123 &  20:58:58.0 & $+$01:22:22.2 & \nodata                   &  $0.32 \pm 0.02$ & $ 76.9\pm 8.8$ &  8.3 & 0.361 &  \\
ACT-CL~J0045.2$-$0152 &  00:45:12.5 & $-$01:52:31.6 &   0.545 \tablenotemark{5} &  $0.55 \pm 0.02$ & $ 53.5\pm 7.3^\ast$ &  7.5 & 0.182 &  \\
ACT-CL~J2050.7$+$0123 &  20:50:43.1 & $+$01:23:29.2 &   0.333 \tablenotemark{5,6} &$0.35 \pm 0.03$ & $ 56.5\pm 7.5$ &  7.4 & 0.104 & RXC J2050.7+0123\tablenotemark{6} \\
ACT-CL~J2128.4$+$0135 &  21:28:23.4 & $+$01:35:36.4 &   0.385 \tablenotemark{5} &  $0.39 \pm 0.03$ & $ 87.2\pm 9.3$ &  7.3 & 0.165 &  \\
ACT-CL~J2025.2$+$0030 &  20:25:12.7 & $+$00:31:33.8 & \nodata                   &  $0.34 \pm 0.02$ & $ 56.6\pm 7.5$ &  6.4 & 0.235 &  \\
ACT-CL~J0026.2$+$0120 &  00:26:15.9 & $+$01:20:37.0 & \nodata                   &  $0.65 \pm 0.04$ & $ 33.6\pm 5.8^\ast$ &  6.3 & 0.196 &  \\
ACT-CL~J2307.6$+$0130 &  23:07:39.9 & $+$01:30:55.8 & \nodata                   &  $0.36 \pm 0.02$ & $ 75.5\pm 8.7$ &  6.1 & 0.027 & ZwCl 2305.0+0114 \tablenotemark{7} \\
ACT-CL~J2156.1$+$0123 &  21:56:08.5 & $+$01:23:27.3 &   0.224 \tablenotemark{4} &  $0.21 \pm 0.02$ & $ 64.9\pm 8.1$ &  6.0 & 0.094 & Abell 2397 \tablenotemark{2} \\
ACT-CL~J0301.6$+$0155 &  03:01:38.2 & $+$01:55:14.6 &   0.167 \tablenotemark{6} &  $0.19 \pm 0.01$ & $ 49.4\pm 7.0$ &  5.8 & 0.069 & RXC J0301.6+0155 \tablenotemark{6} \\
ACT-CL~J2051.1$+$0215 &  20:51:12.2 & $+$02:15:58.3 &   0.321 \tablenotemark{6} &  $0.35 \pm 0.02$ & $ 56.4\pm 7.5$ &  5.2 & 0.221 & RXC J2051.1+0216 \tablenotemark{6} \\
ACT-CL~J0303.3$+$0155 &  03:03:21.1 & $+$01:55:34.5 &   0.153 \tablenotemark{4} &  $0.17 \pm 0.01$ & $ 26.1\pm 5.1$ &  5.2 & 0.060 & Abell 0409 \tablenotemark{2} \\
ACT-CL~J0156.4$-$0123 &  01:56:24.3 & $-$01:23:17.3 & \nodata                   &  $0.45 \pm 0.04$ & $ 37.9\pm 6.2^\ast$ &  5.2 & 0.011 &  \\
ACT-CL~J0219.9$+$0129 &  02:19:52.1 & $+$01:29:52.2 & \nodata                   &  $0.35 \pm 0.02$ & $ 66.0\pm 8.1$ &  4.9 & 0.154 &  \\
ACT-CL~J0240.0$+$0116 &  02:40:01.7 & $+$01:16:06.4 & \nodata                   &  $0.62 \pm 0.03$ & $ 31.9\pm 5.7^\ast$ &  4.8 & 0.077 &  \\
ACT-CL~J0008.1$+$0201 &  00:08:10.4 & $+$02:01:12.3 & \nodata                   &  $0.36 \pm 0.04$ & $ 44.8\pm 6.7$ &  4.7 & 0.028 &  \\
ACT-CL~J0139.3$-$0128 &  01:39:16.7 & $-$01:28:45.2 & \nodata                   &  $0.70 \pm 0.03$ & $ 26.9\pm 5.2^\ast$ &  4.3 & 0.549 &  \\
\enddata
\label{tab:DR8sample}
\tablecomments{R.A.\ and Dec.\ positions denote the BCG location in
  the optical images of the cluster from our confirmation
  procedure. The SZ position was used to construct the ACT descriptor
  identifiers. Spectroscopic redshifts are reported when available and
  come from the DR8 spectroscopic database and our own follow-up with
  GMOS on Gemini South. Values of S/N are from
  \cite{Hasselfield2012}.}

\tablenotetext{1}{from \cite{Gralla2011}}
\tablenotetext{2}{from \cite{Abell1958}}
\tablenotetext{3}{Spectroscopic redshift from \cite{Sarazin1982}}
\tablenotetext{4}{Spectroscopic redshift from \cite{Struble1991}}
\tablenotetext{5}{Spectroscopic redshift from GMOS/Gemini (Sif\'on et al., in prep)}
\tablenotetext{6}{Spectroscopic redshift from \cite{RX}}
\tablenotetext{7}{from \cite{Zwicky1963}}
\tablenotetext{$\ast$}{Denotes clusters at $z>0.4$ for which the $0.4L^*$ limit was not reached and hence richness values are underestimated.}
\end{deluxetable*}

\subsection{Notable Clusters}

In the following sections we provide detailed information on a
selected few individual clusters that are worthy of special attention.

\subsubsection{\JZZFF}
\label{sec:J0044}

\begin{figure}
\centerline{\includegraphics[width=3.4in]{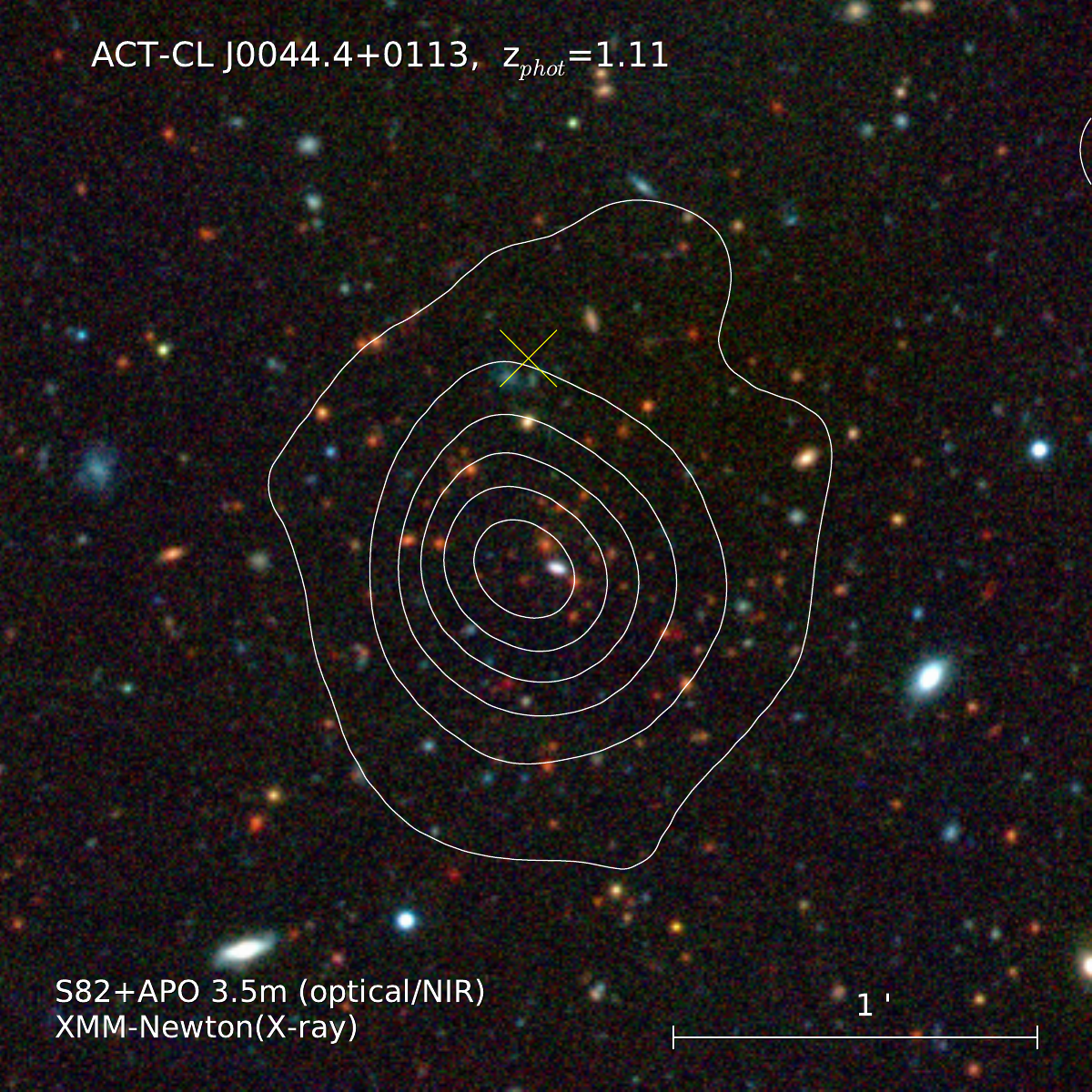}}
\caption{The {\em XMM-Newton} archival X-ray observations (contours)
  that serendipitously contain \JZZFF\ overlaid on the composite
  optical (S82) and NIR (APO) pseudo-color image.  The X-ray
  observations show its extended nature with a best fit
  temperature of $kT_X=7.9\pm1.0$~keV for the photometric
  redshift of $z = 1.11$.}
\label{fig:J0044}
\end{figure}

\JZZFF\ appeared serendipitously in an archival
\xmm\ observation targeting the SLAC lens object SDSSJ0044+0113
\citep{Auger2009} taken on Jan 10, 2010 (PI:Treu, ObsID:0602340101).
After flare rejection we obtained exposure times of 21\,ks for each MOS 
and 15\,ks for the pn.  Our analysis used SAS version 12.0.1.
In Figure~\ref{fig:J0044} we show the composite optical/NIR color
image for \JZZFF\ with the overplotted \xmm\ X-ray surface brightness
contours in the 0.5-4.5 keV band shown in white.  The cluster is
clearly extended and the X-ray surface brightness is above background
up to a radius of $\sim$50$^{\prime\prime}$ (439 $h_{70}^{-1}$ kpc).
Fits to the integrated spectrum to $R_{500c}$ from a region of radius
$1\!'.5$, using a local annular region (covering 2.1$^{\prime}$ to
4.2$^{\prime}$) results in a best fit gas temperature of
$kT_X=7.9\pm1.0$~keV and 0.5--2.0 keV band luminosity of
$L_X=(4.2\pm0.15)\times10^{44}\,h_{70}^{-2}$\,erg~s$^{-1}$, which
assumes the cluster's photometric redshift of $z=1.11$.

We use the \cite{Arnaud2005} $M_{500c}-T_X$ scaling relation based on
\xmm\ observations to estimate the mass for the cluster,
\begin{equation}
M_{500c,T_X} = \frac{M_0}{E(z)}\left(\frac{T_X}{5\,\rm{keV}}\right)^\alpha \, h_{70}^{-1}M_{\odot}
\end{equation}
with $M_0=(3.84\pm0.14)\times10^{14}$, $\alpha=1.71\pm0.09$. The
measured cluster temperature yields a mass of
$M_{500c,T_X}=(4.7\pm1.1)\times10^{14}\,h_{70}^{-1}M_\odot$.
This mass is converted to the mass with respect to the average
density, $M_{200a} = 8.2_{-2.5}^{+3.3}
\times10^{14}\,h_{70}^{-1}M_\odot$ after scaling from critical to
average density using $M_{200a} = 1.77_{-0.17}^{+0.26} \times
M_{500c}$.
This conversion factor was derived using a NFW mass profile and the
concentration-mass relation, $c(M,z)$, from simulations
\citep{Duffy2008} at $z=1.11$ for the mass of the cluster. The
reported uncertainties in the conversion factor reflect the
$\sigma_{{\rm log}\,c}=0.15$ scatter in the log-normal probability
distribution of $c(M,z)$.

The X-ray temperature and inferred mass estimates make \JZZFF\ a
remarkable system that is among the most massive and X-ray-hot
clusters known beyond $z=1$. The mass and temperature of
\JZZFF\ are comparable to the X-ray-discovered cluster
XMMU~J2235.3$-$2557 ($z=1.39$) with $kT_X=8.6\pm1.3$~keV and $M_{200a}=
(8.23\pm1.21) \times 10^{14}\,h_{70}^{-1}M_\odot$
\citep{Rosati2009,Jee2009} and two recent SZ-discovered clusters:
SPT-CL~J2106$-$5844 ($z=1.13$) with $kT_X = 11.0^{+2.6}_{-1.9}$~keV and
$M_{200a}=(1.27\pm0.21)\times10^{15}\,h_{70}^{-1}\,M_\odot$ \citep{Foley2011}
and SPT-CL~J0205-5829 ($z=1.32$) with $kT_X = 8.7^{+1.0}_{-0.8}$~keV
and $M_{500c} =(4.9\pm0.8)\times10^{14}\,h_{70}^{-1}M_\odot$ \citep{Stalder2012}.

\subsubsection{\J2327\ (RCS2 2327)}
\label{sec:J2327}

\begin{figure}
\centerline{\includegraphics[width=3.4in]{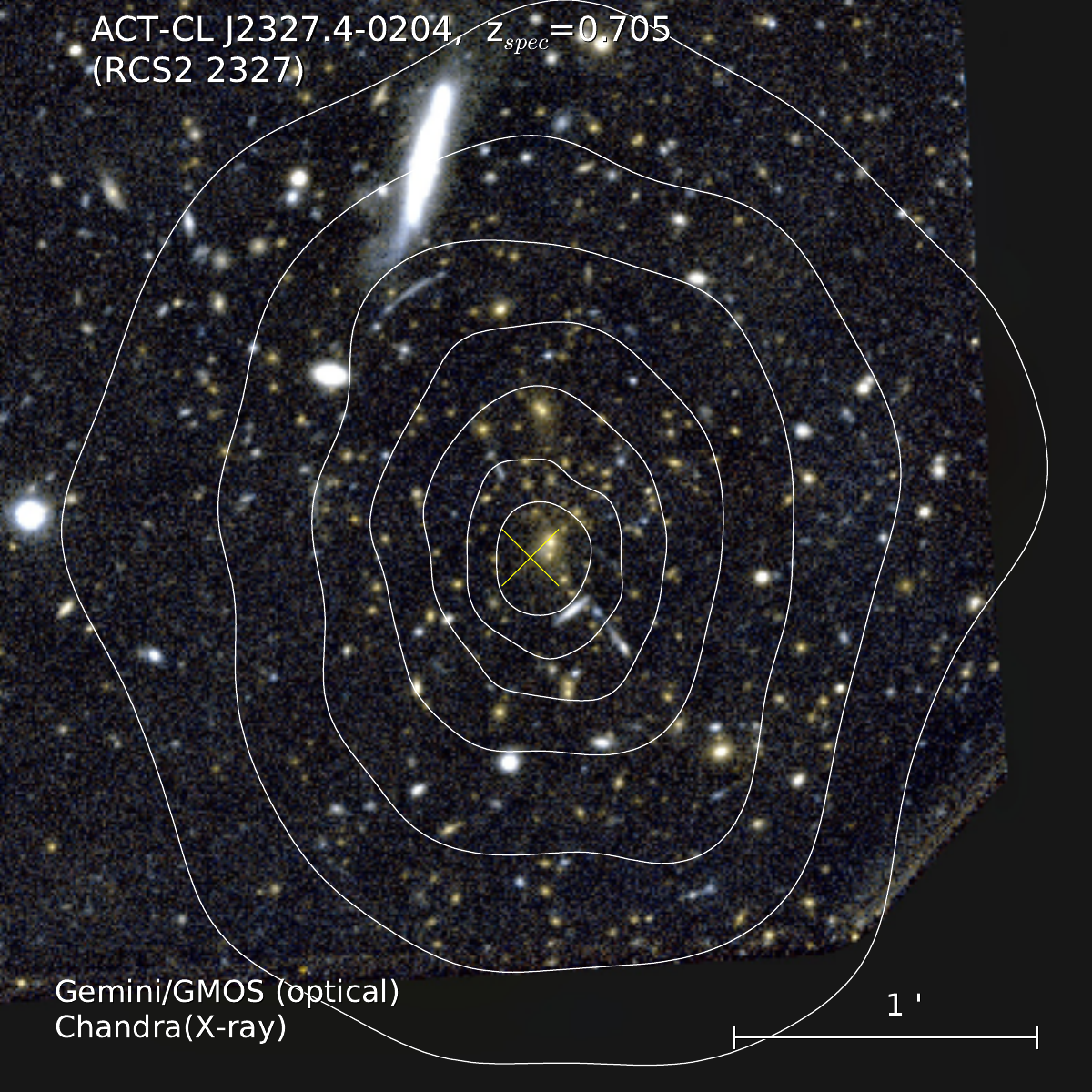}}
\caption{The {\em Chandra} X-ray observation for \J2327, which shows
  its extended nature, is overlaid as white contours on the $gr$
  optical pseudo-color composite image from GMOS. The X-ray contours
  cover a dynamic range of a factor of 100 from the peak to the
  minimum.}
\label{fig:J2327}
\end{figure}

\J2327\ is the cluster with the highest significance detection and the
strongest SZ signal in the full ACT equatorial sample. The cluster has
also been reported as RCS2~2327 by \cite{Gralla2011}. Although the
cluster is not in S82, the system is rich and bright
enough to be detected on the shallower DR8 imaging from which we
obtained an accurate photometric redshift estimate of $z=0.69\pm0.04$
and optical richness $N_{\rm gal}=59.2\pm7.7$.
We searched for archival data and found imaging and spectroscopy from
Gemini/GMOS and X-ray observations from \chandra\ and \xmm. 
We processed the single GMOS pointing (offset $2^\prime$ from the cluster
center) in $g$ ($4\times300$\,s) and $r$ ($4\times300$\,s) taken
on UT 2007 Aug 7 and UT 2007 Dec 26 (GS-2007B-Q-5, PI:Gladders) using
our GMOS custom pipeline \citep{Sifon2012} to create astrometrically
corrected co-added images. The GMOS imaging of the central region of the
cluster, shown in Figure~\ref{fig:J2327}, confirms the picture from DR8
that \J2327\ is a very rich cluster, and reveals the presence of
several strong lensing features.
We also processed the spectroscopic data from the single mask
available taken with the B600 grism for a total integration time of
14.4\,ks of which we were able to process 7.2\,ks. Unfortunately the
setup of the spectroscopic observations only covers the
$4000-6800$\AA\ wavelength range hence putting the CaII K-H absorption
doublet (rest-frame $\lambda_0=3950$\AA) used to secure the
redshift of early-type galaxies at the limit of the detector.
Nevertheless, we were able to extract redshifts for three cluster
galaxies (two of them with [OII] emission), for which we obtain a mean
redshift of $z=0.705$.

A 25\,ks \chandra\ observation (PI:Gladders, ObsID 7355) was taken in
August of 2008 using the ACIS-S array in VFAINT mode. We processed the
data using CIAO version 4.4, applying the latest calibrations (CALDB
version 4.5.0).  VFAINT background rejection was implemented.  X-ray
point sources were identified and compared to the locations of their
optical counterparts, which established that the absolute astrometry
of this \chandra\ observation was good ($1''$).  Background was
subtracted using the blank-sky background files supplied by the
CXC. The process included applying an appropriate filter to the source
data to remove time intervals of high background. This observation was
devoid of any background flares.

For making images, point sources were removed and replaced with Poisson
distributed counts based on the surrounding level of background or
source emission. Exposure maps were created in the soft (0.5-2 keV)
band.  Figure~\ref{fig:J2327} shows surface brightness contours of the
background-subtracted, exposure-corrected, adaptively-smoothed
\chandra\ X-ray data in the 0.5--2 keV band. The X-rays show a
strongly peaked distribution centered very close to the BCG.  The
cluster X-ray isophotes are modestly elliptical with an axial ratio of
$\sim$1.2 and little centroid shift.  We detect X-ray emission out to
a mean radius of $\sim0.\!'8$ ($670\,h^{-1}_{70}$\,kpc).

\begin{figure}
\centerline{
\includegraphics[width=3.6in]{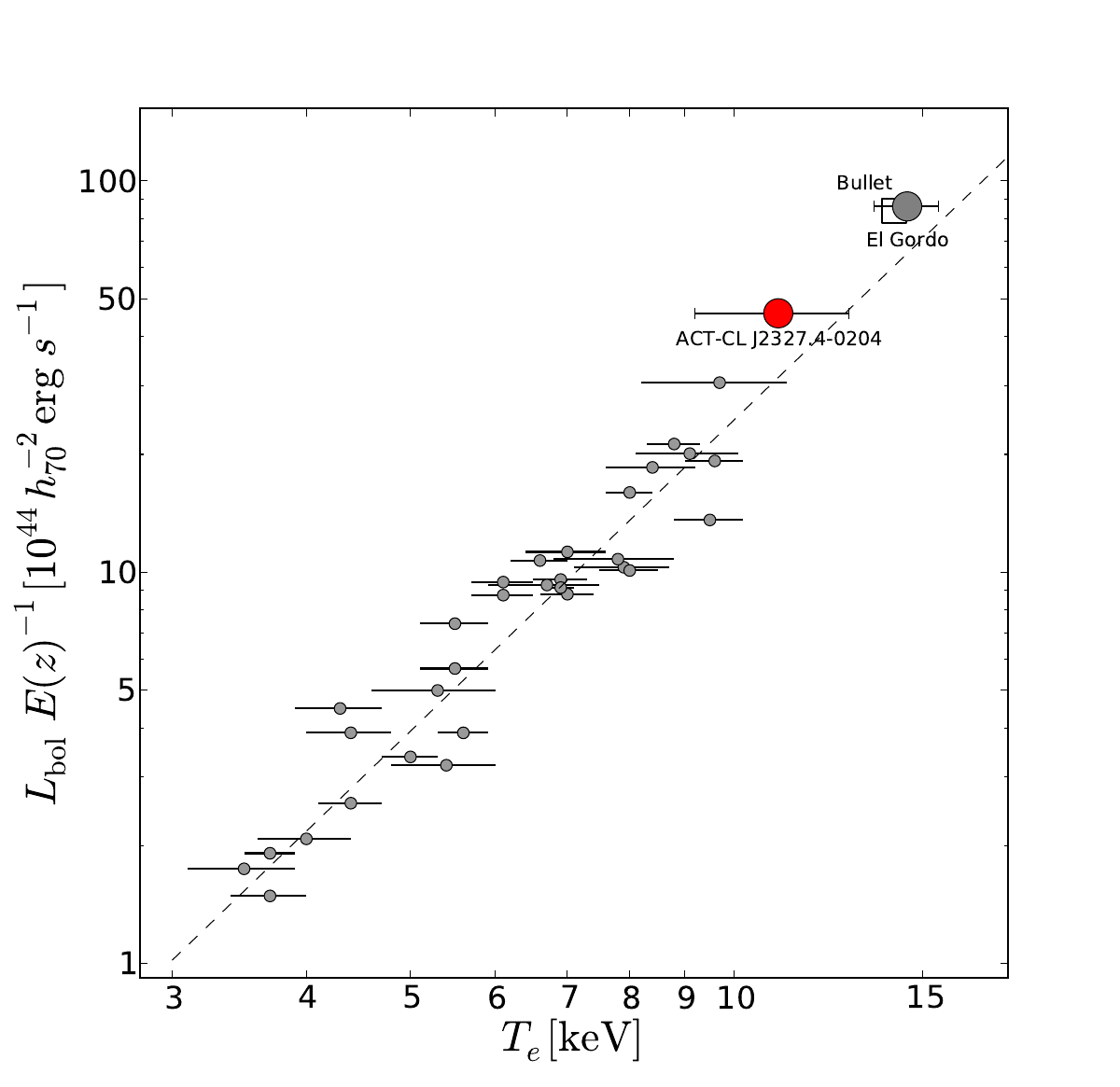}
}
\caption{The X-ray bolometric luminosity vs.\ temperature for a sample
  of well-studied clusters taken from \cite{Markevitch1998}. The
  Bullet cluster (1E0657$-$56) and El Gordo are the open square point
  and gray circle at high temperature and luminosity from
  \cite{Markevitch2006} and \cite{Menanteau2012}
  respectively. \J2327\ is the red circle that shows the remarkable
  properties of the cluster. The dashed line represents the $L-T$
  best-fit from \cite{Markevitch1998}.}
\label{fig:L-T}
\end{figure}

We use this observation to measure the gas temperature of the
cluster. An absorbed {\tt phabs*mekal} model yielded a best-fit
(source frame) temperature of $kT_X = 11.0_{-1.3}^{+2.0}$~keV from the
core-excised \chandra\ spectrum (covering from $0.15 R_{500c}$ to
$0.50R_{500c}$ after iterating to obtain $R_{500c}$). We use this
value with Eq.~5 in \cite{Vikhlinin2009} to estimate the cluster
temperature within $R_{500c}$ which can then be used in the
$T_X$-$M_X$ scaling law.  We also have obtained an integrated spectrum
(covering the full cluster out to a radius of $\sim0.\!'8$.  From this
we obtain a bolometric luminosity of $L_{\rm
  bol}=(6.6\pm0.3)\times10^{45}\, h_{70}^{-2}$~erg~s$^{-1}$.
Figure~\ref{fig:L-T} shows the $L_{\rm bol}$-$T_X$ relation with
\J2327 added as the red point. The smaller grey points show the sample
of \cite{Markevitch1998}, while the white square and large grey circle
show the closest comparison clusters, 1E0657$-$56 and El
Gordo. Similarly, the X-ray luminosity of \J2327 in the 0.5--2.0 keV
band is $L_X=1.39\pm0.05\times10^{45}\,h_{70}^{-2}$erg~s$^{-1}$.

We follow the prescriptions in \cite{Vikhlinin2009} and apply the
$T_X-M_{500c}$ scaling law to the \chandra\ data and obtain a mass of
$M_{500c,T_X} = 9.7^{+3.1}_{-1.8}\times 10^{14}h_{70}^{-1}M_\odot$.
We also investigated the cluster mass from $M_{\rm gas}$ using the
scaling law for $M_{500c}-M_{\rm gas}$ at redshift $z=0.6$ from
\cite{KVN}, which yields a value of $M_{500c,\rm Mgas} =
(9.6\pm1.2)\times10^{14}h_{70}^{-1}M_\odot$ and implies a gas mass
fraction $f_{\rm gas} = 0.12$. Both X-ray derived masses are in good
agreement and confirm the view of an exceptional and massive cluster.
We use the weighted-average of the $T_X$ and $M_{\rm gas}$ mass
estimates to obtain a combined mass for \J2327\ of $M_{500c}
=(1.0\pm0.1)\times10^{15}\,h_{70}^{-1}M_\odot$. We convert the combined
mass with respect to the average density using the same procedure as
for ACT-CL~J0044.4$+$0113 in Section~\ref{sec:J0044} using the scaling
$M_{200a}=1.86_{-0.20}^{+0.30}\times M_{500c}$ which produces a mass
$M_{200a}=1.9_{-0.4}^{+0.6}\times10^{15}\,h_{70}^{-1}M_\odot$. 
This mass is also consistent with the velocity dispersion of
$\approx1400$\,km~s$^{-1}$ presented by \citet{Yee2009}\footnote{http://malaysia09.nottingham.ac.uk/} which
indicates a dynamical mass of
$M_{200a,\rm dyn}\approx2.6\times10^{15}\,h_{70}^{-1}M_\odot$ when
applying the same procedure as described below (Section~\ref{sec:J0022}).
In summary, the high SZ-signal, gas temperature, gas mass and velocity
dispersion, taken together, establish \J2327\ as one of the two most
massive clusters known at $z>0.6$, the other being El Gordo.

\subsubsection{ACT-CL~J0022.2$-$0036}
\label{sec:J0022}

ACT-CL~J0022.2$-$0036 is the highest significance SZ detection on the
S82 region of the ACT equatorial sample and has been extensively
targeted for our follow-up observations. \cite{Reese2012} recently
presented SZA observations of the cluster, while \cite{Miyatake2012}
estimated its mass from weak-lensing measurements using Subaru. As
part of our spectroscopic follow-up with Gemini (Sif\'on et al. in prep)
we have secured redshifts for 44 members from which we obtain a
redshift $z=0.8054 \pm 0.0013$ and velocity dispersion of $\sigma_{\rm
  gal}=1213 \pm 155$~\kms. We use the $M_{200c}-\sigma_{\rm DM}$
scaling relation from \cite{Evrard2008} to convert the measured
velocity dispersion into a dynamical mass estimate,
\begin{equation}\label{eq:mass}
 M_{200c} = \frac{10^{15}}{0.7 E(z)}\left(\frac{\sigma_{\rm
     DM}}{\sigma_{{\rm DM}, 15}}\right)^{1/{\alpha}}\,h_{70}^{-1}\,M_\odot,
\end{equation}
where $\sigma_{{\rm DM},15} = 1082.9\pm4.0$~km~s$^{-1}$, $\alpha =
0.3361\pm0.0026$ and $\sigma_{\rm DM}$ is the velocity dispersion of
the dark matter halo. The latter is related to the observed galaxy
velocity dispersion by the velocity bias parameter, $b_v = \sigma_{\rm
  gal}/\sigma_{\rm DM}$. The latest physically motivated simulations
\citep[see][and references therein]{Evrard2008} indicate that galaxies
are essentially unbiased tracers of the dark matter potential,
$\langle b_v \rangle=1.00\pm0.05$.
Using a bias factor of $b_v=1$ for the velocity dispersion for all
galaxies, we obtain a dynamical mass of $M_{200a,\rm dyn}=
1.5_{-0.6}^{+0.7}\times 10^{15}\,h_{70}^{-1}\, M_\odot$,  using
the conversion factor $M_{200a}=1.17_{-0.03}^{+0.04} M_{200c}$ with the same
prescription as described in Section~\ref{sec:J0044}.
This value is consistent with the Subaru weak-lensing mass of
$M_{200a, {\rm WL}}=1.1^{+0.6}_{-0.5}\times 10^{15}\,h_{70}^{-1}\,
M_\odot$ recently reported by \cite{Miyatake2012}.

\section{Discussion}

\subsection{The Purity of the S82 Sample}
\label{sec:purity}

\begin{figure}
\centerline{\includegraphics[width=3.9in]{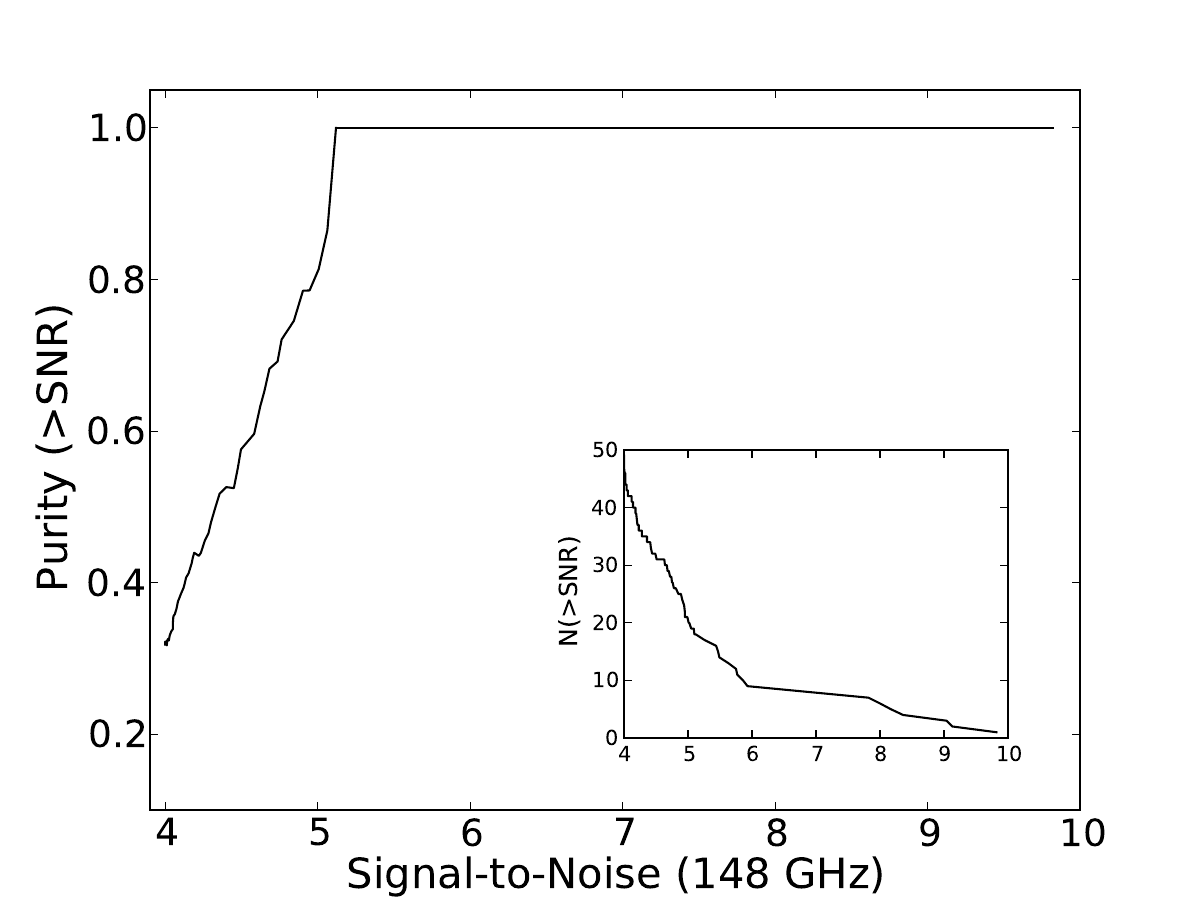}}
\caption{The S82 SZ cluster sample purity greater than a given S/N in
  the ACT 148 GHz maps. The purity is defined as the ratio of the
  number of confirmed clusters to the number of observed clusters
  candidates. The solid line represents the purity binned in $n=2$
  events. The inset plot shows the cluster cumulative distribution as
  a function of S/N for the optically-confirmed cluster sample.}
\label{fig:purity}
\end{figure}

The purity for an SZ sample is defined as the ratio of
optically-confirmed clusters to SZ detections \citep{Menanteau-SZ}.
In Figure~\ref{fig:purity} we show the purity from the sample of 155
SZ cluster candidates with S/N$>4.0$ within the S82 region as a
function of signal-to-noise. A notable improvement from our previous
work on the southern sample \citep{Menanteau-SZ} is that for the S82
region we were able to examine {\em every} cluster candidate
regardless of its signal-to-noise. The inset plot shows the cluster
cumulative distribution as a function of S/N for the
optically-confirmed cluster sample.
We achieve 100\% purity for signal-to-noise ratios greater than 5.1
where there are 19 clusters. This drops to a purity value of 80\% for
a SNR of 5.0.  
For SZ candidates down to a signal-to-noise of 4.6 the sample the
purity is 60\% (31/52).
Below this S/N value we find a purity value of only 30\% down to a
signal-to-noise of 4.0.
It is important to mention that we did not perform targeted follow-up
in the NIR for all SZ candidates without a clear optical
identification. We obtained $K_S$ for all of candidates above
S/N$>5.1$, but only for a fraction of those above S/N$>4.7$. Therefore
some of the clusters that were not optically confirmed could potentially be
at $z>1$.
The purity of the S82 sample is consistent with the purity we found
for the ACT southern sample.

\subsection{Cluster X-ray Properties}
\label{sec:xray}

\begin{figure*}
\centerline{\includegraphics[width=7in]{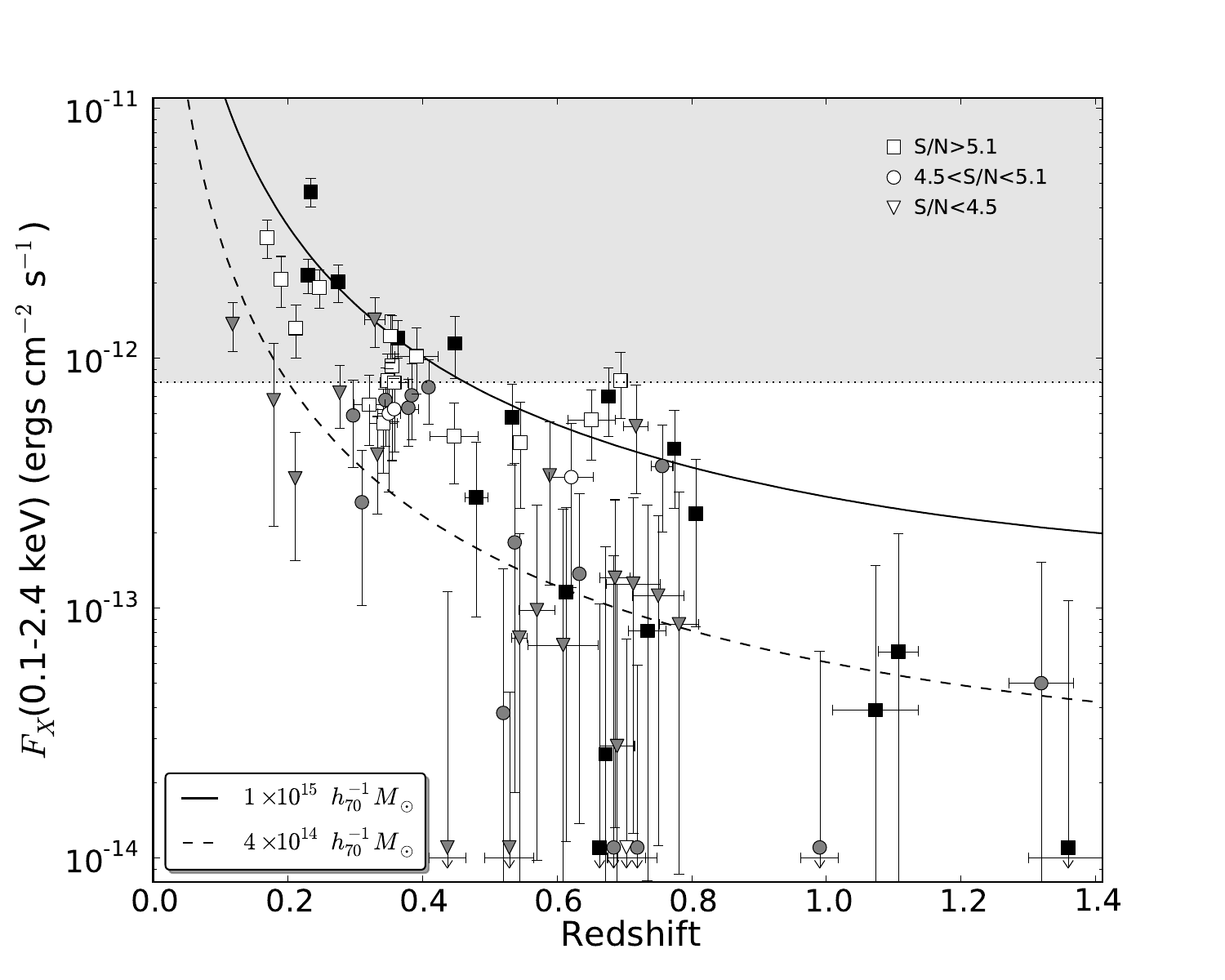}}
\caption{The soft X-ray flux (in the observed band of 0.1--2.4~keV)
  from RASS archival data as a function of redshift for ACT SZ
  equatorial clusters. Clusters with S/N$>5.1$, are shown as boxes,
  circles represent clusters with $4.5<$S/N$<5.1$ and triangles
  clusters S/N$<4.5$. White filled symbols represent objects outside
  the S82 the deep region. For non detection sources we show upper
  limits. The dashed and solid curves represent the expected X-ray
  fluxes for clusters with masses
  $M_{200a}=4\times10^{14}\,h_{70}^{-1}M_\odot $ and
  $1\times10^{15}\,h_{70}^{-1}M_\odot$, respectively, using scaling
  relations from \cite{Vikhlinin2009}. The gray area at the top
  corresponds approximately to the regime of the RASS bright source
  catalog \citep{Voges-99}.}
\label{fig:fx-z}
\end{figure*}

For consistency with our previous work \citep{Menanteau-SZ} and to
establish qualitatively that the ACT clusters indeed comprise a
massive sample, we present the RASS X-ray fluxes and luminosities for all
confirmed ACT clusters, regardless of the significance of the RASS
detection in Tables~\ref{tab:S82RASS} and \ref{tab:DR8RASS}.
The soft RASS X-ray fluxes as a function of redshift are plotted in
Figure~\ref{fig:fx-z}.  We indicate the region at the high-end flux
that approximately corresponds to the flux limit of the RASS bright
source catalog \citep{Voges-99}. In cases of no significant detection
we indicate upper limits. Clusters detected with higher significance
(S/N$>4.5$) in the ACT 148\,GHz data are shown as black squares while
the others are represented by gray circles.
The low statistical quality of the RASS data for most of these
clusters precludes making accurate estimates of cluster masses from
the X-ray data. However, for reference we show curves of the expected
(observed-frame) X-ray fluxes for clusters with assumed masses of
$M_{200a}=4\times10^{14}h_{70}^{-1}M_\odot $ (dashed) and
$1\times10^{15}h_{70}^{-1} M_\odot $ (solid) using the X-ray
luminosity versus mass scaling relation in \cite{Vikhlinin2009}.  We
use their Eq.~22, which includes an empirically determined redshift
evolution.  We convert their X-ray band (emitted: 0.5--2 keV band) to
ours (observed: 0.1--2.4 keV) assuming a thermal spectrum at the
estimated cluster temperature determined using the mass-temperature
relation also from \cite{Vikhlinin2009}.  This too has a redshift
dependence, so the estimated temperatures vary in the ranges 2.9--5.0
keV and 5.2--9.0 keV over $0.0<z<1.1$ for the two mass values we plot
(solid and dashed lines).  We use conversion factors assuming the
redshift-averaged temperatures of 4 keV and 7 keV, since the
difference in conversion factor over the temperature ranges is only a
few percent, negligible on the scale of Figure~\ref{fig:fx-z}.  The
mass values in \cite{Vikhlinin2009} are defined with respect to an
overdensity of 500 times the {\it critical} density of the Universe at
the cluster redshift. As we have done before, we convert to an
overdensity of 200 times the {\it average} density following the
procedure in Section~\ref{sec:J0044}. This mass conversion factor is
approximately 1.8 averaged over redshift, varies from 2.2 to 1.7 over
$0.0<z<1.1$, and depends only weakly on cluster mass (few percent) for
the 2 values plotted here.

The X-ray fluxes for the ACT SZ-selected clusters scatter around the
$4\times10^{14}\,h_{70}^{-1}M_\odot$ and
$1\times10^{15}\,h_{70}^{-1}M_\odot$ curves, for the full and pure
samples respectively, validating that this is a massive cluster sample.

\subsection{Mass Estimation}
\label{sec:masses}

\begin{figure}
\centerline{\includegraphics[width=4.0in]{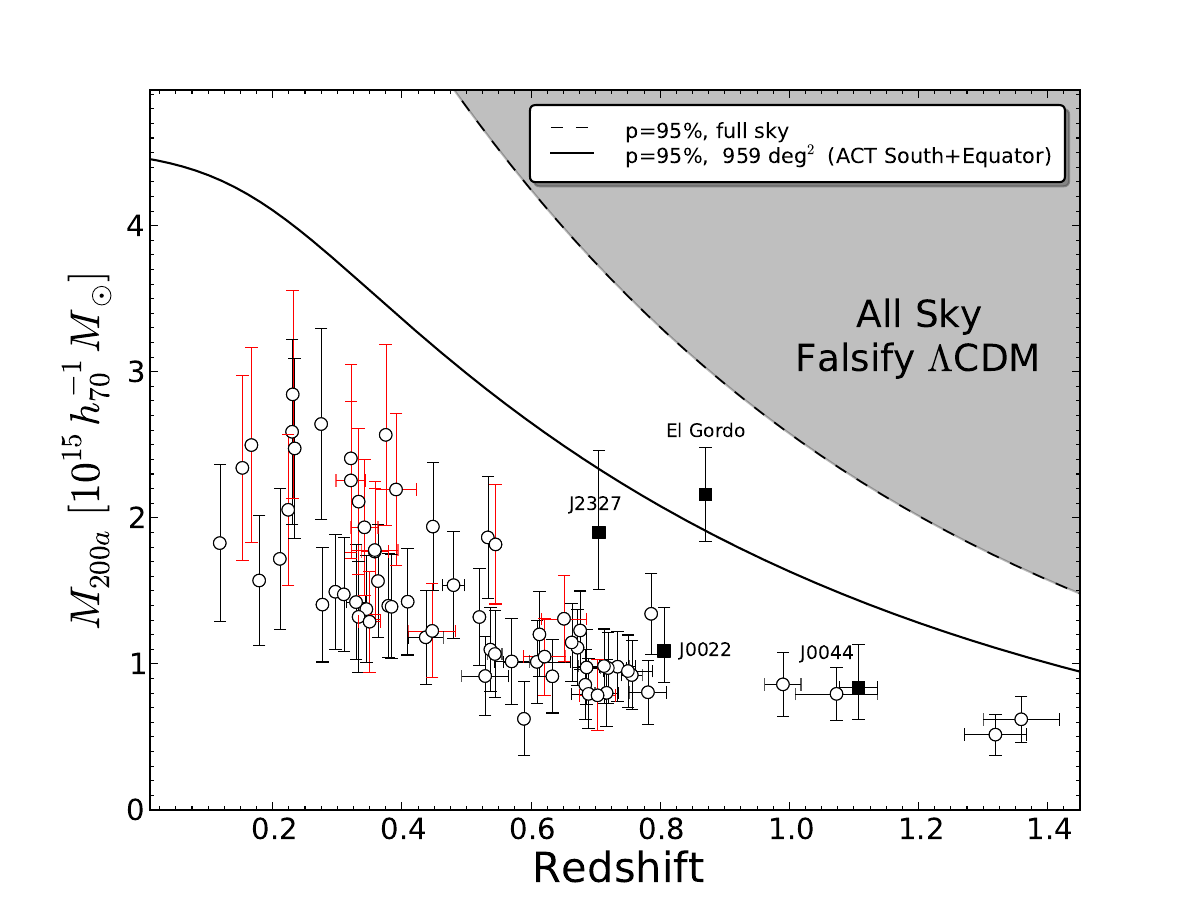}}
\caption{The exclusion curves, $M_{200a}(z)$, from the
  \cite{Mortonson2011} fitting formulas where a single cluster above
  the corresponding curve would conflict with Flat $\Lambda$CDM and
  quintessence at 95\% C.L for both sample and parameters
  variance. The dashed and solid lines represent the exclusion curves
  for the full sky and the ACT survey analyzed region of 959~deg$^2$
  (i.e. the mass limit for a cluster at a given redshift that it is
  less than 5\% likely to be found in 95\% of the $\Lambda$CDM
  parameter probability distribution). The open circles with black and
  red error bars represent the $M_{200a}$ mass estimates using the
  $\widetilde{y}_0-M$ scaling law \citep{Sifon2012} for clusters inside and
  outside of S82 respectively. We also show the X-ray mass
  estimates based for \JZZFF\ and \J2327\ (combined) as black squares. For
  ACT-CL~J0022.2$-$0036 we show its dynamical mass from the Gemini
  observations. For reference, we also show our combined mass estimate
  for El Gordo, with 1-$\sigma$ error bars \citep{Menanteau2012}.}
\label{fig:Mass-z}
\end{figure}

In order to obtain consistent estimates across the full ACT equatorial
sample we use the $\widetilde{y}_0-M_{200c}$ scaling relation from
\cite{Sifon2012}, which relates the fixed aperture central Compton
decrement $\widetilde{y}_0$ as described by \cite{Hasselfield2012} and
$M_{200c}$ (critical) for the cluster as
\begin{equation}
M_{200c} = 10^{(15.03 \pm 0.07)} \,\left(\frac{\widetilde{y}_0 E(z)^{-2}}{5\times10^{-5}}\right) ^ {0.74\pm0.11}  h_{70}^{-1} M_\odot .
\end{equation}
We use the $\widetilde{y}_0$ measurements from Table~3 from
\cite{Hasselfield2012} to estimate $M_{200c}$ for the sample which in
turn we transform to $M_{200a}$ (average) using the same procedure
described in Section~\ref{sec:J0044}. 
For a comparison with the dynamical masses from \cite{Sifon2012} and the
effects of different models of gas pressure profile in cluster SZ mass
see \cite{Hasselfield2012}. 

In Figure~\ref{fig:Mass-z} we show the SZ-derived masses for the whole
sample as a function of their redshift using the above mass-scaling relation.
We also compare the masses of the cluster sample with the
$M_{200a}(z)$ exclusion curves from \cite{Mortonson2011} for which a
single cluster with mass $M_{200a}$ above the corresponding curve
would conflict with $\Lambda$CDM and quintessence at 95\% confidence
level, including both sample and cosmological parameter variance. In
other words, the exclusion curves represent the mass threshold as a
function of redshift for which any cluster is less than 5\% likely to be
found in a survey region for 95\% of the $\Lambda$CDM parameter
variance.
In order to address the rarity of any of our clusters in
Fig.~\ref{fig:Mass-z} we plot the exclusion curves for the full sky
and the region analyzed for the ACT survey (959~deg$^2$).  
We note, however, that \cite{Harrison2012} have recently suggested
that these curves tend to overestimate the amount of tension with
cosmological models because they underestimate the number of massive
clusters due to an inadequate {\em a posteriori} choice of mass and
redshift cuts.
In the figure, we show the X-ray mass estimates for \JZZFF\ and \J2327
(combined), plus the dynamical mass of ACT$-$CL~J0022.2$-$0036 and El
Gordo (for comparison), as black squares.

\subsection{Distance between SZ Centroid and BCG}

An important source of systematic uncertainty in studies of large
optically-selected clusters is the ``miscentering'' of the BCG with
respect to the dark matter (DM) and hot gas. This may be the result
from either a misidentification of the BCG by the finding algorithm,
or a real physical offset between the DM and gas centers and the
BCG. The latter can be a {\em real} astrophysical observable linked to
the cluster dynamical state, while the former is essentially just a
flaw in the identification algorithm \citep[see][for
  examples of recent studies]{Skibba2011,George2012}.

\begin{figure}
\centerline{\includegraphics[width=4.0in]{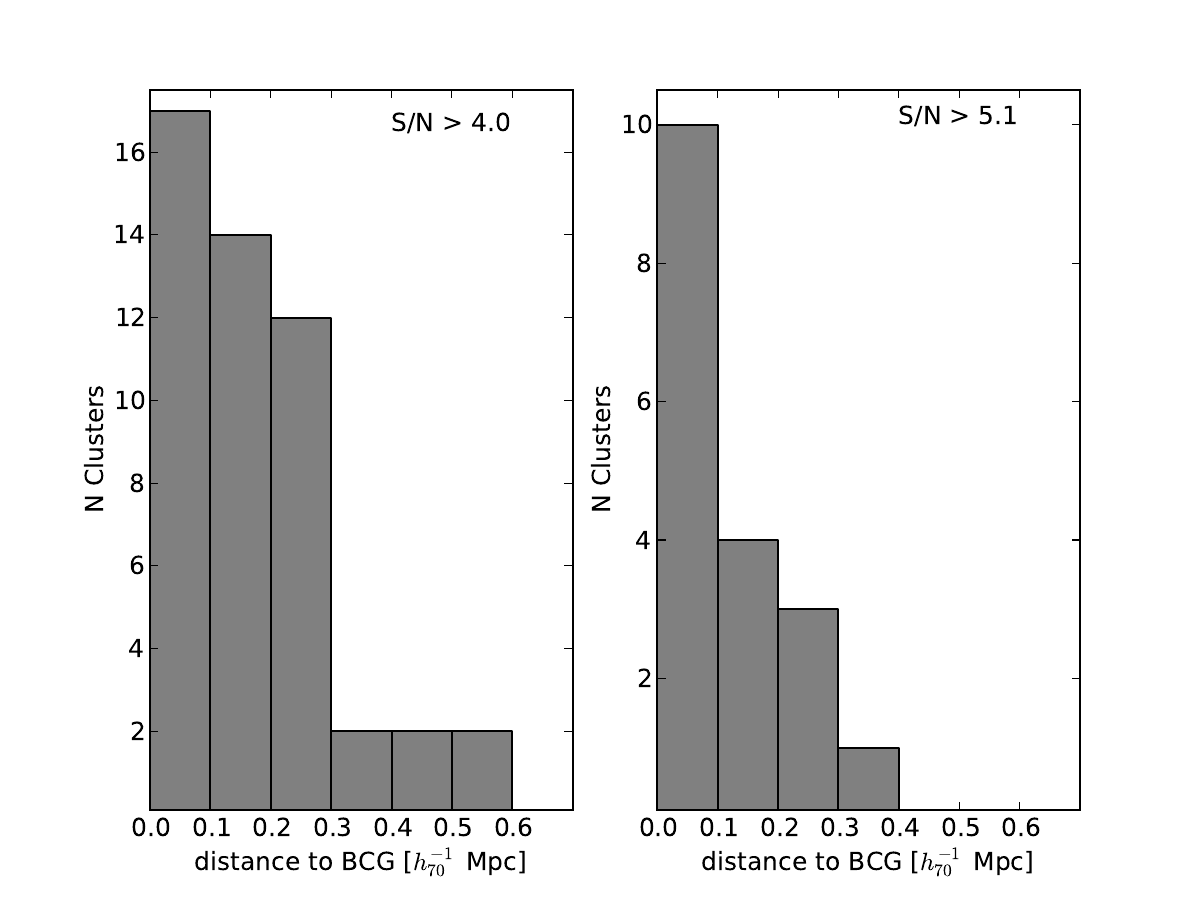}}
\caption{The distribution of the distances in co-moving units between
  the ACT centroid SZ position and the location of the BCG in the optical
  images for the high-significance sample (right panel, S/N$>5.1$) and
  the full sample (left panel, S/N$>4$).}
\label{fig:d2BCG}
\end{figure}

Prompted by the results from \cite{Planck-scaling} where the predicted
$Y_{SZ}$ from the $Y_{SZ}-N_{200}$ relation \citep{Rozo2009} was much
higher than the observed stacked $Y_{SZ}$ signal for the MaxBCG
clusters \citep{MaxBCG}, some studies have considered miscentering to
explain this discrepancy \citep[e.g.][]{Biesiadzinski2012,Sehgal2012},
while others favor solutions that include uncertainties in optical
systematics and selection effects \citep{Angulo2012,Rozo2012}.
Our sample of SZ galaxy clusters allows us to directly investigate the
offsets between the SZ centroid position in the 148\,GHz ACT maps and
the location of the BCG on the optical images. One key advantage of
our sample is that the BCG identification has been performed visually
in all cases to ensure that it is indeed the brightest member in the
cluster and hence is virtually free of bias due to miscentering.
We compute the offsets between the BCG and SZ centroid position for
the high-significance sample (S/N$>5.1$) and the full sample
(Figure~\ref{fig:d2BCG}) for the S82 region. We find that the typical
distance to the BCG is less than $0.3$\,Mpc\,$h_{70}^{-1}$ for both
samples, with mean distance values of $0.12$\,Mpc\,$h_{70}^{-1}$ and
$0.17$\,Mpc\,$h_{70}^{-1}$ for the high-significance and the full
sample respectively.
We therefore find no significant evidence for the amount of
misalignment required to explain the discrepancy between $Y_{SZ}$ and
$N_{200}$ by {\em Planck} to be a physical offset, as opposed to
algorithmic, between the BCG and the gas center.
We note that the typical positional uncertainty of the ACT SZ centroid
for cluster is $\approx1'.4/$ SNR which for S/N$=4.5$ corresponds to
$0.12\,h_{70}^{-1}$\,Mpc and $0.16\,h_{70}^{-1}$\,Mpc at $z=0.5$ and
$z=1.0$ respectively. Thus the observed level of offset can be
accounted by the uncertainty in the ACT cluster centering.

\subsection{SZ-Optical Richness Relation}

\begin{figure}
\centerline{\includegraphics[width=4.0in]{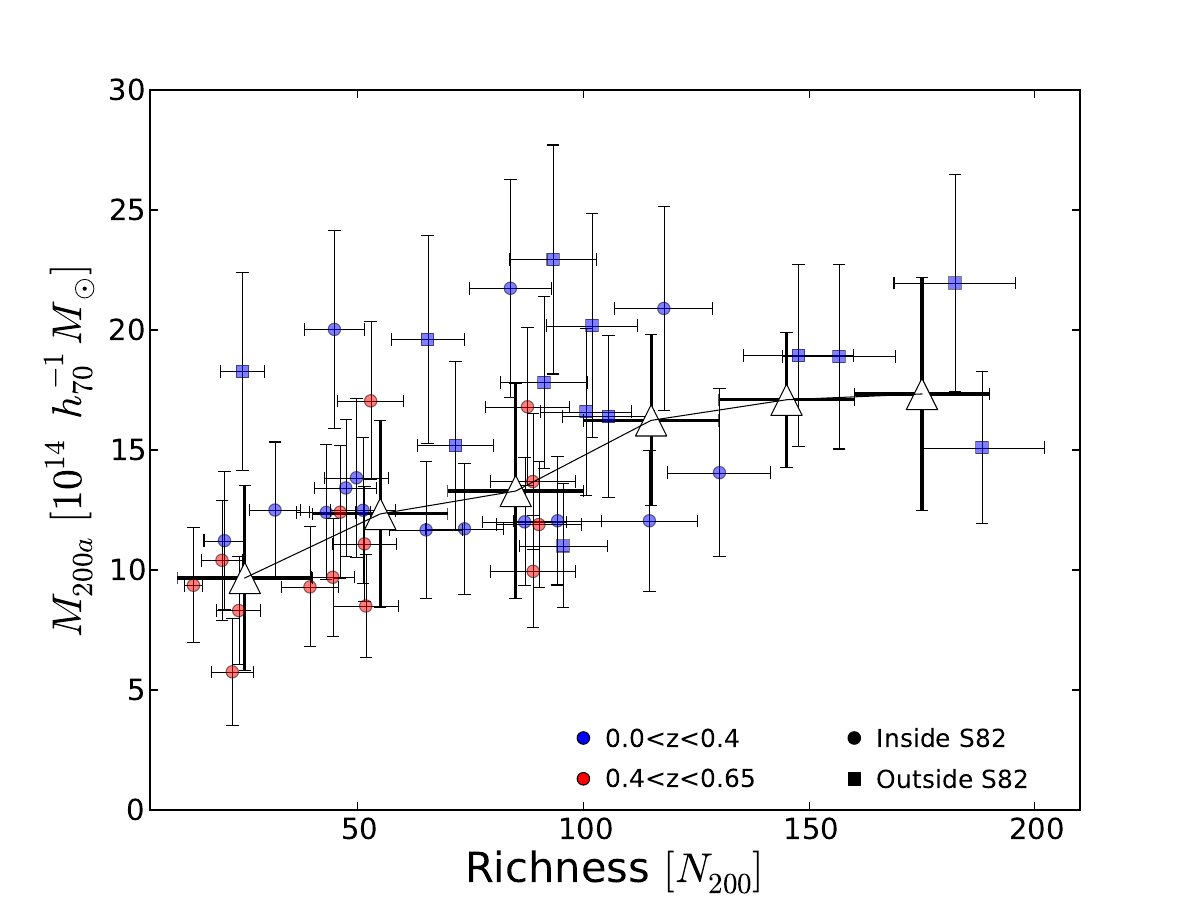}}
\caption{The SZ-derived $M_{200a}$ mass from using \cite{Sifon2012}
  $\widetilde{y_0}-M$ scaling relation for SZ equatorial clusters
  compared to their optical richness $N_{200}$. We show clusters on
  S82 as circles and systems outside that area as
  squares. Additionally, we colored red the systems located at $z>0.4$
  and blue at $z<0.4$. Open triangles denote the weighted mean mass
  value in bins of size $\Delta N_{200}=30$.}
\label{fig:N200}
\end{figure}

The continuous optical coverage provided by SDSS over S82 and DR8 for
our cluster sample also enables a direct and independent investigation
of the relation between optical richness and SZ signal.
For this purpose, we use our own computed $N_{200}$ values for
clusters as described in Section~\ref{sec:membership} and compare with
their SZ-derived $M_{200a}$ masses (using the same values as above in
Section~4.3).  We only consider clusters at $z<0.65$ within S82 and
$z<0.40$ outside S82 to take into account the different flux limits of
the two samples to ensure unbiased values of $N_{200}$ and hence allow
for a meaningful comparison.
In Figure~\ref{fig:N200} we show the results with symbols keyed to the
location of the clusters (inside or outside S82) color-keyed to their
redshifts.
A simple examination of the individual points hints of a weak
correlation with high scatter between the optical richness with SZ
mass. For this data set we calculate a Pearson's correlation
coefficient of $r=0.52$ and probability of an uncorrelated system
producing a correlation at least as extreme as the one computed of
$p=0.0004$ which indicates a real correlation but with significant scatter.
In order to bring visual clarity in the scatter we estimate the
mean-weighted SZ mass as a function of $N_{200}$ in bins of size $\Delta
N_{200}=30$ which we also show in Figure~\ref{fig:N200}. The errors
were estimated as the quadratic sum of the weighted errors and the
error in the mean.
We conclude that optical richness estimates such as $N_{200}$ for
modest-sized samples like this one do not provide precise mass
estimates.

\section{Summary and Conclusions}

We present the optical/NIR confirmation and physical properties of a
new sample of 68 SZ detected clusters from ACT over the celestial
equator.
Our study takes advantage of the wide and deep coverage over S82 which
with additional pointed NIR observations enabled the characterization
of 49 clusters up to $z\approx1.3$. Although this is a well-studied
region of the sky, 22 of the 49 clusters on the S82 region are newly
discovered systems all lying at $z>0.55$, highlighting the power of
the SZ effect to discover massive clusters at high redshift. Moreover,
five of these clusters are at $z\geq1$.
Outside the S82 region we use the regular-depth SDSS data from DR8
to confirm 19 additional clusters to $z\approx0.7$, with 10
systems discovered by ACT.
We have also analysed ROSAT All-Sky Survey data for the sample, which
confirms that this is a high-mass cluster sample.

The S82 optical data provides a powerful complement as it allows the
study of every SZ cluster candidate down to any ACT S/N
desired. Preliminary inspection of lower S/N candidates finds several
rich optical systems, suggesting that improved multi-wavelength
cluster-finding algorithms may allow for additional discoveries.

We investigate differences in the location of the BCG and the SZ
centroid positions and we find no evidence for significant offsets
between them. We also study the relation between the optical richness
and cluster mass and find only a weak correlation between both
quantities.

As with the ACT southern sample, we find some spectacular systems in
the ACT equatorial sample. We report on the discovery of \JZZFF, at
$z=1.1$, with mass and X-ray temperature that put it in the league of
the extreme clusters at $z>1$. We also present a multi-wavelength
analysis of the rich cluster \J2327 at $z=0.705$. This is the cluster
with the highest SZ significance in the whole ACT equatorial sample and
is comparable to systems like El Gordo and the Bullet
Cluster.

Based on currently available mass estimates there is no tension between
\lcdm\ and the sample's mass-redshift distribution. El Gordo and
\J2327\ are the most extreme clusters in the joint ACT sample, and
more detailed, multi-wavelength follow-up studies will aid in further
constraining the masses and physical properties of these clusters, as
well as their cosmological implications.

\begin{deluxetable*}{clcrrrr}
\tablecaption{\rosat\ Properties of Stripe 82 ACT equatorial cluster} 
\tablewidth{0pt}
\tablehead{
  \colhead{} & 
  \colhead{} & 
  \colhead{$t_{\rm exp}$} &
  \colhead{$N_H$} &
  \colhead{$R$} &
  \colhead{$F_X$\tablenotemark{a}} &
  \colhead{$L_X$\tablenotemark{b}}
  \\
  \colhead{ACT Descriptor} & 
  \colhead{$z$} & 
  \colhead{(s)} & 
  \colhead{($10^{20}$ cm$^{-2}$)} &
  \colhead{($h_{70}^{-1}$ kpc)} &
  \colhead{(0.1$-$2.4 keV)} &
  \colhead{(0.1$-$2.4 keV)}
}
\startdata
ACT-CL~J0012.0$-$0046 & 1.36  &  369 &  3.22 &  1514 & $ 0.01\pm0.97$ & $ 0.11\pm7.59$ \\
ACT-CL~J0014.9$-$0057 & 0.533 &  357 &  3.06 &  1136 & $ 5.81\pm2.07$ & $ 5.30\pm1.89$ \\
ACT-CL~J0017.6$-$0051 & 0.211 &  351 &  2.93 &   619 & $ 3.30\pm1.75$ & $ 0.39\pm0.21$ \\
ACT-CL~J0018.2$-$0022 & 0.75  &  372 &  2.69 &  1321 & $ 1.12\pm1.22$ & $ 2.23\pm2.43$ \\
ACT-CL~J0022.2$-$0036 & 0.806 &  398 &  2.76 &  1355 & $ 2.38\pm1.54$ & $ 5.59\pm3.61$ \\
ACT-CL~J0044.4$+$0113 & 1.11  &  304 &  1.86 &  1473 & $ 0.67\pm1.31$ & $ 3.29\pm6.39$ \\
ACT-CL~J0051.1$+$0055 & 0.69  &  365 &  2.41 &  1278 & $ 0.28\pm0.96$ & $ 0.46\pm1.58$ \\
ACT-CL~J0058.0$+$0030 & 0.76  &  384 &  2.82 &  1324 & $ 3.70\pm1.68$ & $ 7.50\pm3.40$ \\
ACT-CL~J0059.1$-$0049 & 0.77  &  356 &  3.33 &  1336 & $ 4.34\pm1.84$ & $ 9.29\pm3.92$ \\
ACT-CL~J0104.8$+$0002 & 0.277 &  428 &  3.31 &   758 & $ 7.27\pm2.05$ & $ 1.54\pm0.43$ \\
ACT-CL~J0119.9$+$0055 & 0.72  &  425 &  3.12 &  1300 & $<0.49$ & $<0.89$ \\
ACT-CL~J0127.2$+$0020 & 0.379 &  431 &  2.90 &   935 & $ 6.33\pm1.90$ & $ 2.69\pm0.81$ \\
ACT-CL~J0152.7$+$0100 & 0.230 &  432 &  2.75 &   661 & $21.49\pm3.33$ & $ 3.04\pm0.47$ \\
ACT-CL~J0206.2$-$0114 & 0.68  &  388 &  2.53 &  1268 & $ 7.02\pm2.15$ & $11.01\pm3.37$ \\
ACT-CL~J0215.4$+$0030 & 0.73  &  164 &  2.83 &  1310 & $ 0.81\pm1.77$ & $ 1.53\pm3.34$ \\
ACT-CL~J0218.2$-$0041 & 0.672 &  194 &  2.96 &  1265 & $ 0.26\pm1.50$ & $ 0.40\pm2.32$ \\
ACT-CL~J0219.8$+$0022 & 0.537 &  182 &  2.88 &  1140 & $ 1.83\pm1.95$ & $ 1.70\pm1.81$ \\
ACT-CL~J0221.5$-$0012 & 0.589 &  211 &  2.75 &  1193 & $ 3.39\pm2.16$ & $ 3.88\pm2.48$ \\
ACT-CL~J0223.1$-$0056 & 0.663 &  220 &  2.94 &  1258 & $<0.94$ & $<1.41$ \\
ACT-CL~J0228.5$+$0030 & 0.72  &  229 &  2.28 &  1298 & $ 5.34\pm2.48$ & $ 9.58\pm4.46$ \\
ACT-CL~J0230.9$-$0024 & 0.44  &  190 &  2.07 &  1019 & $<1.06$ & $<0.62$ \\
ACT-CL~J0241.2$-$0018 & 0.684 &  192 &  2.95 &  1274 & $<1.52$ & $<2.44$ \\
ACT-CL~J0245.8$-$0042 & 0.179 &   89 &  3.43 &   544 & $ 6.78\pm4.66$ & $ 0.56\pm0.38$ \\
ACT-CL~J0250.1$+$0008 & 0.78  &  183 &  5.09 &  1340 & $ 0.86\pm2.06$ & $ 1.88\pm4.49$ \\
ACT-CL~J0256.5$+$0006 & 0.363 &  689 &  6.16 &   910 & $12.06\pm2.07$ & $ 4.65\pm0.80$ \\
ACT-CL~J0301.1$-$0110 & 0.53  & 1281 &  6.92 &  1131 & $<0.36$ & $<0.32$ \\
ACT-CL~J0308.1$+$0103 & 0.633 &  295 &  5.93 &  1233 & $ 1.37\pm1.50$ & $ 1.85\pm2.03$ \\
ACT-CL~J0320.4$+$0032 & 0.384 &  351 &  6.31 &   943 & $ 7.10\pm2.38$ & $ 3.10\pm1.04$ \\
ACT-CL~J0326.8$-$0043 & 0.448 &  300 &  6.76 &  1034 & $11.46\pm3.20$ & $ 7.07\pm1.97$ \\
ACT-CL~J0336.9$-$0110 & 1.32  &  511 &  7.61 &  1510 & $ 0.50\pm1.02$ & $ 3.61\pm7.40$ \\
ACT-CL~J0342.0$+$0105 & 1.07  &  365 &  7.44 &  1464 & $ 0.39\pm1.09$ & $ 1.79\pm4.96$ \\
ACT-CL~J0342.7$-$0017 & 0.310 &  385 &  6.04 &   820 & $ 2.65\pm1.63$ & $ 0.72\pm0.44$ \\
ACT-CL~J0348.6$-$0028 & 0.345 &  357 &  9.87 &   881 & $ 6.80\pm2.36$ & $ 2.34\pm0.81$ \\
ACT-CL~J0348.6$+$0029 & 0.297 &  344 &  9.96 &   796 & $ 5.90\pm2.25$ & $ 1.46\pm0.56$ \\
ACT-CL~J2050.5$-$0055 & 0.613 &  432 &  6.24 &  1215 & $ 1.16\pm1.37$ & $ 1.46\pm1.72$ \\
ACT-CL~J2051.1$+$0056 & 0.333 &  460 &  7.44 &   860 & $ 4.11\pm1.73$ & $ 1.31\pm0.55$ \\
ACT-CL~J2055.4$+$0105 & 0.409 &  456 &  7.54 &   980 & $ 7.65\pm2.20$ & $ 3.85\pm1.11$ \\
ACT-CL~J2129.6$+$0005 & 0.234 &  275 &  3.63 &   670 & $46.33\pm6.05$ & $ 6.81\pm0.89$ \\
ACT-CL~J2130.1$+$0045 & 0.71  &  277 &  3.71 &  1295 & $ 1.25\pm1.51$ & $ 2.22\pm2.68$ \\
ACT-CL~J2135.1$-$0102 & 0.33  &  323 &  3.87 &   853 & $14.24\pm3.24$ & $ 4.42\pm1.01$ \\
ACT-CL~J2135.7$+$0009 & 0.118 &  362 &  4.27 &   384 & $13.69\pm3.04$ & $ 0.47\pm0.10$ \\
ACT-CL~J2152.9$-$0114 & 0.69  &  331 &  7.46 &  1276 & $ 1.32\pm1.39$ & $ 2.14\pm2.26$ \\
ACT-CL~J2154.5$-$0049 & 0.48  &  352 &  7.42 &  1075 & $ 2.77\pm1.85$ & $ 2.00\pm1.33$ \\
ACT-CL~J2220.7$-$0042 & 0.57  &  231 &  4.74 &  1174 & $ 0.98\pm1.61$ & $ 1.04\pm1.72$ \\
ACT-CL~J2229.2$-$0004 & 0.61  &  243 &  4.73 &  1211 & $ 0.71\pm1.77$ & $ 0.88\pm2.19$ \\
ACT-CL~J2253.3$-$0031 & 0.54  &  356 &  5.43 &  1148 & $ 0.76\pm1.23$ & $ 0.73\pm1.18$ \\
ACT-CL~J2302.5$+$0002 & 0.520 &  348 &  4.02 &  1121 & $ 0.38\pm1.05$ & $ 0.32\pm0.91$ \\
ACT-CL~J2337.6$+$0016 & 0.275 &  379 &  3.55 &   754 & $20.21\pm3.46$ & $ 4.23\pm0.72$ \\
ACT-CL~J2351.7$+$0009 & 0.99  &  376 &  3.48 &  1438 & $<0.57$ & $<2.13$ \\
\enddata
\tablenotetext{}{$N_H$  represents the Galactic column density of
  neutral hydrogen at the cluster location and $R$ the physical radius
 corresponding to 3$^\prime$ at the cluster's redshift.}
\tablenotetext{a}{Units are $\times 10^{-13}$ erg s$^{-1}$ cm$^{-2}$.}
\tablenotetext{b}{Units are $\times 10^{44}$ erg s$^{-1}$.}
\label{tab:S82RASS}
\end{deluxetable*}

\begin{deluxetable*}{clcrrrr}
\tablecaption{\rosat\ Properties of ACT equatorial cluster outside Stripe 82} 
\tablewidth{0pt}
\tablehead{
  \colhead{} & 
  \colhead{} & 
  \colhead{$t_{\rm exp}$} &
  \colhead{$N_H$} &
  \colhead{$R$} &
  \colhead{$F_X$\tablenotemark{a}} &
  \colhead{$L_X$\tablenotemark{b}}
  \\
  \colhead{ACT Descriptor} & 
  \colhead{$z$} & 
  \colhead{(s)} & 
  \colhead{($10^{20}$ cm$^{-2}$)} &
  \colhead{($h_{70}^{-1}$ kpc)} &
  \colhead{(0.1$-$2.4 keV)} &
  \colhead{(0.1$-$2.4 keV)}
}
\startdata
ACT-CL~J0008.1$+$0201 & 0.36  &  386 &  2.72 &   902 & $ 6.25\pm2.04$ & $ 2.34\pm0.76$ \\
ACT-CL~J0026.2$+$0120 & 0.65  &  504 &  2.86 &  1248 & $ 5.67\pm1.77$ & $ 8.16\pm2.54$ \\
ACT-CL~J0045.2$-$0152 & 0.545 &  311 &  2.81 &  1149 & $ 4.58\pm2.07$ & $ 4.40\pm1.99$ \\
ACT-CL~J0139.3$-$0128 & 0.70  &  320 &  2.89 &  1288 & $<0.65$ & $<1.11$ \\
ACT-CL~J0156.4$-$0123 & 0.45  &  424 &  2.56 &  1033 & $ 4.87\pm1.72$ & $ 2.99\pm1.06$ \\
ACT-CL~J0219.9$+$0129 & 0.35  &  164 &  3.31 &   889 & $ 6.00\pm3.08$ & $ 2.13\pm1.09$ \\
ACT-CL~J0239.8$-$0134 & 0.375 &   84 &  3.01 &   897 & $ 9.32\pm5.43$ & $ 3.42\pm1.99$ \\
ACT-CL~J0240.0$+$0116 & 0.62  &  235 &  2.92 &  1222 & $ 3.34\pm2.13$ & $ 4.32\pm2.75$ \\
ACT-CL~J0301.6$+$0155 & 0.167 &  214 &  6.57 &   570 & $20.72\pm4.82$ & $ 1.94\pm0.45$ \\
ACT-CL~J0303.3$+$0155 & 0.153 &  254 &  6.61 &   519 & $30.35\pm5.30$ & $ 2.21\pm0.39$ \\
ACT-CL~J2025.2$+$0030 & 0.34  &  444 &  9.53 &   876 & $ 5.49\pm2.04$ & $ 1.85\pm0.69$ \\
ACT-CL~J2050.7$+$0123 & 0.333 &  461 &  7.71 &   886 & $ 8.12\pm2.24$ & $ 2.85\pm0.79$ \\
ACT-CL~J2051.1$+$0215 & 0.321 &  462 &  7.87 &   892 & $12.26\pm2.66$ & $ 4.42\pm0.96$ \\
ACT-CL~J2058.8$+$0123 & 0.32  &  454 &  6.73 &   840 & $ 6.52\pm2.05$ & $ 1.91\pm0.60$ \\
ACT-CL~J2128.4$+$0135 & 0.39  &  281 &  4.07 &   954 & $10.17\pm3.00$ & $ 4.63\pm1.36$ \\
ACT-CL~J2135.2$+$0125 & 0.231 &  398 &  4.25 &   697 & $19.15\pm3.36$ & $ 3.17\pm0.56$ \\
ACT-CL~J2156.1$+$0123 & 0.224 &  317 &  4.79 &   621 & $13.20\pm3.17$ & $ 1.57\pm0.38$ \\
ACT-CL~J2307.6$+$0130 & 0.36  &  341 &  4.31 &   902 & $ 7.99\pm2.41$ & $ 2.99\pm0.90$ \\
ACT-CL~J2327.4$-$0204 & 0.705 &  367 &  4.74 &  1282 & $ 8.14\pm2.41$ & $13.56\pm4.02$ \\
\enddata
\tablenotetext{}{$N_H$  represents the Galactic column density of
  neutral hydrogen at the cluster location and $R$ the physical radius
 corresponding to 3$^\prime$ at the cluster's redshift.}
\tablenotetext{a}{Units are $\times 10^{-13}$ erg s$^{-1}$ cm$^{-2}$.}
\tablenotetext{b}{Units are $\times 10^{44}$ erg s$^{-1}$.}
\label{tab:DR8RASS}
\end{deluxetable*}

\acknowledgements

This work was supported by the U.S. National Science Foundation
through awards AST-0408698 and AST-0965625 for the ACT project, and
PHY-0855887, PHY-1214379 and AST-0707731.  Funding was also provided
by Princeton University, the University of Pennsylvania, a Canada
Foundation for Innovation (CFI) award to UBC, and CONICYT awards to
PUC. 
ACT operates in the Parque Astron\'omico Atacama in northern Chile under
the auspices of the Comisi\'on Nacional de Investigaci\'on
Cient\'ifica y Tecnol\'ogica de Chile (CONICYT).
\chandra\ and \xmm\ X-ray studies on ACT clusters at Rutgers are
supported by Chandra grants GO1-12008X, GO1-13156X and NASA ADAP grant
NNX11AJ48G, respectively.
Computations were performed on the GPC supercomputer at the SciNet HPC
Consortium.  SciNet is funded by the CFI under the auspices of Compute
Canada, the Government of Ontario, the Ontario Research Fund --
Research Excellence; and the University of Toronto.

Based in part on observations obtained with the Apache Point
Observatory 3.5-meter telescope, which is owned and operated by the
Astrophysical Research Consortium.

Funding for SDSS-III has been provided by the Alfred P. Sloan
Foundation, the Participating Institutions, the National Science
Foundation, and the U.S. Department of Energy Office of Science. The
SDSS-III web site is http://www.sdss3.org/.
SDSS-III is managed by the Astrophysical Research Consortium for the
Participating Institutions of the SDSS-III Collaboration including the
University of Arizona, the Brazilian Participation Group, Brookhaven
National Laboratory, University of Cambridge, Carnegie Mellon
University, University of Florida, the French Participation Group, the
German Participation Group, Harvard University, the Instituto de
Astrofisica de Canarias, the Michigan State/Notre Dame/JINA
Participation Group, Johns Hopkins University, Lawrence Berkeley
National Laboratory, Max Planck Institute for Astrophysics, Max Planck
Institute for Extraterrestrial Physics, New Mexico State University,
New York University, Ohio State University, Pennsylvania State
University, University of Portsmouth, Princeton University, the
Spanish Participation Group, University of Tokyo, University of Utah,
Vanderbilt University, University of Virginia, University of
Washington, and Yale University.

\end{document}